\documentclass[twocolumn, numberedappendix]{aastex62}

\usepackage{amstext}
\usepackage{amsmath}
\usepackage{amssymb}
\usepackage{graphicx}

\newcommand{\beq}{\begin{equation}}
\newcommand{\eeq}{\end{equation}}
\newcommand{\beqar}{\begin{align}}
\newcommand{\eeqar}{\end{align}}

\shorttitle{}

\shortauthors{}

\begin{document}

\title{Bound Outflows, Unbound Ejecta, and the Shaping of Bipolar Remnants during Stellar Coalescence}

\author[0000-0002-1417-8024]{Morgan MacLeod}
\altaffiliation{NASA Einstein Fellow}
\affiliation{Harvard-Smithsonian Center for Astrophysics, 60 Garden Street, Cambridge, MA, 02138, USA}
\email{morgan.macleod@cfa.harvard.edu}

\author[0000-0002-0509-9113]{Eve C. Ostriker}
\affiliation{Department of Astrophysical Sciences, Princeton University, Princeton, NJ 08544, USA}

\author[0000-0001-5603-1832]{James M. Stone}
\affiliation{Department of Astrophysical Sciences, Princeton University, Princeton, NJ 08544, USA}

\begin{abstract}
Recent observations have revealed that the remnants of stellar-coalescence transients are bipolar. This raises the questions of how these bipolar morphologies arise and what they teach us about the mechanisms of mass ejection during stellar mergers and common-envelope phases. In this paper, we analyze hydrodynamic simulations of the lead-in to binary coalescence, a phase of unstable Roche lobe overflow that takes the binary from the Roche limit separation to the engulfment of the more compact accretor within the envelope of the extended donor.  As mass transfer runs away at increasing rates, gas trails away from the binary. Contrary to previous expectations, early mass loss from the system remains bound to the binary and forms a circumbinary torus. Later ejecta, generated as the accretor grazes the surface of the donor, have very different morphologies and are unbound. These two components of mass loss from the binary interact as later, higher-velocity ejecta collide with the circumbinary torus formed by earlier mass loss. Unbound ejecta are redirected toward the poles, and escaping material creates a bipolar outflow. Our findings show that the transition from bound to unbound ejecta from coalescing binaries can explain the bipolar nature of their remnants, with implications for our understanding of the origin of bipolar remnants of stellar-coalescence transients  and, perhaps, some preplanetary nebulae. 
\end{abstract}


\section{Introduction}

Luminous red novae are optically-discovered transients, many of which are somewhat brighter and redder than classical novae. This class of sources are believed to originate from stellar mergers.  A key object for making this identification was the galactic transient V1309 Sco \citep{2010A&A...516A.108M}, which was a contact, eclipsing binary with a decreasing orbital period prior to undergoing a 10 magnitude outburst \citep{2011A&A...528A.114T}. Following the outburst, any sign of periodicity had vanished, suggesting that the outburst marked the coalescence of the binary system. 

The majority of luminous red novae discovered to date share a primary star evolving up the giant branch that has likely grown in radius to the point that it interacts with its companion.  \citet{1976IAUS...73...75P} described the subsequent evolution as a common envelope phase -- in which one star engulfs its companion and the two stellar cores spiral closer under the influence of drag forces. In the decades since this proposal, the details of such common envelope phases have remained major sources of theoretical uncertainty \citep[][]{1993PASP..105.1373I,2000ARA&A..38..113T,2010NewAR..54...65T,2013A&ARv..21...59I}. 

Among the theoretical uncertainties about common envelope interactions are questions about how mass is ejected and how binary systems map through this phase of transformation. Which systems will emerge as transformed (and tightened) binaries? Which systems will merge into a single remnant? Such questions are of particular importance for understanding the source population for gravitational wave detection experiments like LIGO and VIRGO \citep{2016PhRvD..93l2003A,2016PhRvL.116x1103A,2017PhRvL.118v1101A,2017PhRvL.119p1101A,2017ApJ...851L..35A}. 

The evidence presented by luminous red novae transients offer an avenue for progress. It has become clear that these sources represent emission from ejecta generated during the plunge of a coalescing binary toward the onset of a common envelope phase \citep[e.g.][]{2006MNRAS.373..733S,2013Sci...339..433I, 2017ApJ...835..282M,2017ApJ...850...59P,2017MNRAS.471.3200M}. In studying these transients, we are directly observing the properties of long-uncertain mass ejection during common envelope episodes \citep{2017ApJ...835..282M}.  By studying the remnants of transients over the decades following their outbursts, we can also hope to distinguish the outcome of a given interaction. 

A particularly puzzling mystery has emerged in the morphology of remnants of luminous red nova transients. These remnants show bipolar morphologies, mostly clearly presented in new Atacama Large Millimeter Array (ALMA) observations by \citet{2018arXiv180401610K}, though observations of optical polarization offered an earlier clue \citep{2011A&A...527A..75K,2013A&A...558A..82K}. How are these bipolar morphologies attained? One line of thinking has held that the process is itself bipolar -- as in magnetically-collimated jets that direct ejecta in polar directions \citep[e.g.][]{2012ApJ...746..100S,2016RAA....16...99K,2016MNRAS.462..217S,2017MNRAS.467.3299K,2018MNRAS.480.1898C}, perhaps leading the more compact star to never become completely engulfed in the envelope as a result \citep{2015ApJ...800..114S,2016NewA...47...16S,2017MNRAS.465L..54S,2018MNRAS.477.2584S}.

\citet{2006MNRAS.365....2M,2007Sci...315.1103M,2009MNRAS.399..515M} and, later, \citet{2017MNRAS.471.3200M,2017ApJ...850...59P} have considered the alternative mechanism for generating aspherical morphologies. These authors each considered spherical explosion triggered inside a rotationally-supported torus (presumed to be formed by preceding binary interaction), which results in a collimation of ejecta toward the system's poles.  
\citet{2018ApJ...860...19G}, \citet{2018arXiv180705925F}, \citet{2018arXiv180902297R} have recently discussed the interaction of stellar winds with the debris of a common envelope episode, and how this process might contribute to observed bipolar planetary nebulae. 
Here we expand on this line of work to show that bipolar morphologies can, instead, arise from interaction of successive components of equatorial mass loss from the coalescing binary.

We analyze numerical simulations of coalescing binaries transitioning from Roche lobe overflow toward the onset of the common envelope phase. 
We show that early outflows, contrary to previous theoretical predictions, are bound to the binary system and form a circumbinary torus. Later ejecta, generated just as the secondary object plunges within the envelope of the donor, have larger radial velocities and are unbound. The interaction of unbound ejecta with the bound torus redirects ejecta toward the poles, laying the foundation for the generation of bipolar remnants of stellar coalescence, recently reported by \citet{2018arXiv180401610K}.

\section{Bipolar Remnants of Stellar Coalescence and Binary Interaction}\label{sec:bipolarobsv}

\subsection{Remnants of Galactic Red Novae}
There is growing evidence that the remnants of galactic luminous red novae outbursts, believed to originate from stellar coalescence, can exhibit bipolar structures. There is strong evidence for a bipolar morphology of V4332 Sgr. We also review the less-extensive evidence related to the two other galactic transients that have been studied, V1309 Sco and V838 Mon.  

Some of the first evidence for bipolar morphologies came from V and R band linear polarization measurements of the remnant of V4332 Sgr, which flared in 1994 \citep{2011A&A...527A..75K}. Correcting for the effect of significant unpolarized line emission, \citet{2011A&A...527A..75K} estimate continuum polarizations of approximately 40\% and 20\% for V and R band, respectively, and suggest that dust scattering in a bipolar outflow is one of the only morphologies that can produce such high linear polarization in an unresolved source. Spectropolarimetric observations by \citet{2013A&A...558A..82K} confirmed \citet{2011A&A...527A..75K}'s suggested continuum polarization. 

Earlier conclusions regarding morphology have recently been strengthened by submillimeter observations, which have offered an opportunity to spatially resolve galactic remnants of stellar-coalescence transients \citep{2018arXiv180401610K}. Spatially-resolved observations using the Atacama Large Millimeter Array (ALMA) of V4332 Sgr produce a position-velocity diagram that is best-modeled with a bipolar molecular outflow with a 60 degree opening angle (see Figures 5 and 6 of \citet{2018arXiv180401610K}). 

V1309 Sco's merger has been extremely useful in decoding observations of luminous red novae in general. Data here are highly suggestive of a multi-component outflow, but not conclusively in favor of a bipolar structure. The system is viewed within a few degrees of edge on; it was an eclipsing binary prior to its merger \citep{2011A&A...528A.114T}.
In reporting the outburst, \citet{2010A&A...516A.108M} show broad Balmer-line profiles of emission with approximate Full-width at half maximum (FWHM) of 150~km~s$^{-1}$, with narrow embedded absorption components of roughly 80~km~s$^{-1}$ suggestive of a partially-obscuring, slower component (like an equatorial outflow).
 Relatedly, \citet{2017ApJ...850...59P} have very successfully modeled the changing pre-outburst eclipsing lightcurve reported by \citet{2011A&A...528A.114T} with increasing mass-loss from the vicinity of $L_2$.
V1309 Sco is only marginally spatially resolved in ALMA data presented by \citet{2018arXiv180401610K}. The comparison of optical \citep{2015A&A...580A..34K} and submillimeter (rotational) spectral lines shows ongoing recombination of hydrogen, and evidence for shock-heated gas via [OI] $\lambda 6300$ emission with kinematics related to (but not identical to) the molecular outflow -- perhaps suggestive of a multi-component morphology with continuous interaction similar to V4332 Sgr, though not yet conclusive \citep{2018arXiv180401610K}. 

Finally, as shown by \citet{2014A&A...569L...3C}, mid-infrared emission surrounding V838 Mon is highly extended. It has not yet been spatially resolved, however, so we cannot yet conclude whether this represents an edge-on disk-like structure or a bipolar structure \citep{2018arXiv180401610K}. 

\subsection{The Uncertain Origin of Bipolar Planetary Nebulae}

A larger, and perhaps related, question is that of the ``shapes and shaping of planetary nebulae'' as a review by \citet{2002ARA&A..40..439B} is titled. The formation processes of bipolar and asymmetric planetary nebulae have long been subjects of debate \citep[see, for example, the recent discussion by one such source by][]{2018PASA...35...27M}. Recently, the kinematics of very young, bipolar planetary nebulae have been mapped in detail with ALMA \citep[e.g.][]{2013A&A...557L..11B, 2016A&A...593A..92B, 2017A&A...597L...5B, 2013ApJ...777...92S,2017ApJ...835L..13S}, a Chandra survey has offered a high-energy perspective on planetary nebulae and their central objects \citep{2012AJ....144...58K,2014ApJ...794...99F,2015ApJ...800....8M}, and many new identifications of binary central stars in planetary nebulae have been made \citep[see section 6.3 of ][for a recent review]{2017PASA...34....1D}. 
Within these sources there is extensive evidence for dense, mass rich torii that shape the remaining outflow (likely wind from the remnant central star) into the observed profiles of nebular emission \citep[see, for example, early simulations by][]{1989ApJ...339..268S}. However, the origin of this toroidal distribution of gas is more uncertain, with possible explanations ranging from slow, equatorially concentrated winds of stars at the tip of the asymptotic giant branch to binary interaction \citep[see the reviews of][]{1981ApJ...249..572M,2002ARA&A..40..439B,2017PASA...34....1D,2017NatAs...1E.117J}.

\citet{2018ApJ...860...19G} have recently studied the hydrodynamic shaping of stellar winds into planetary nebulae using 2D simulations initialized with the ejecta properties of a previous common-envelope simulation. \citet{2018arXiv180705925F} report a qualitatively similar calculation in 3D. 
Other theoretical suggestions have focused on the conversion of accretion energy into jets \citep{1994ApJ...421..219S,2012ApJ...746..100S,2013MNRAS.436.1961A,2015ApJ...800..114S,2015MNRAS.453.2115A,2017MNRAS.471.4839S,2018MNRAS.477.2584S,2018MNRAS.480.1898C}. 
As emphasized by \citet{2018arXiv180705925F}, the emergent bipolar signatures of these models are tantalizing potential explanations of planetary nebulae morphologies.

\section{Numerical Method and Simulations}\label{method}

We report on results from simulations of the gas dynamics of binary coalescence created with the {\tt Athena++} hydrodynamics code (Stone, J. M. et al., in preparation).\footnote{version 1.0, url: https://princetonuniversity.github.io/athena} We solve conservation equations of mass, momentum, energy, for inviscid hydrodynamics, with additional source terms that describe the gravitational potential of the binary. We perform our calculation in the frame of the (orbiting) donor, primary star. This aids in the fidelity with which we can model the donor star's hydrostatic equilibrium, and as a result, the early phases of mass transfer, but introduces fictitious accelerations in the equation of motion. We adopt an ideal gas equation of state with adiabatic index $\gamma=5/3$. A full description of the equations solved, source terms, and tests is given in \citet{2018ApJ...863....5M}. 

Our simulated system is a giant-star donor that fills its Roche lobe and transfers mass onto a less-massive, more-compact donor (modeled as a point mass). The mass ratio of accretor to donor is $q=M_2/M_1 = 0.3$. The simulation is performed in dimensionless units in which the donor's original mass and radius and the gravitational constant are all unity. With this choice of units, the time unit in the simulation is the characteristic donor-star dynamical time, $(R_1^3/GM_1)^{1/2}$. The orbital time is $P_{\rm orb}=2\pi \left(a^3 / G (M_1 + M_2)  \right)^{1/2}$, where $a$ is the orbital semi-major axis. Therefore, in code units, $P_{\rm orb} =2\pi \left(a/R_1 \right)^{3/2} \left(1+q\right)^{-1/2}$. 
Simulated results may be rescaled to physical systems from these values. For example, if the simulated donor star were a $1M_\odot$, $10R_\odot$ giant, our characteristic time unit is 0.58~day, the initial orbital period is 9.5~days, the unit velocity is $(GM_1/R_1)^{1/2}=140$~km~s$^{-1}$, and the unit density is $M_1/R_1^3=5.9\times10^{-3}$~g~cm$^{-3}$. Additional examples of this rescaling are given in Table 1 of \citet{2018ApJ...863....5M}. 

Our simulation begins at the Roche limit separation, $2.06R_1$. Mass transfer from the donor to the accretor leads to accelerating orbital decay, and we stop the calculation when the accretor has plunged within the envelope of the donor (at $0.6R_1$).  

We will show results in a coordinate system in which the $x-y$ plane is the orbital plane of the binary, with the angular momentum vector oriented in the $+z$-direction. Inertial-frame coordinates relative to the system center of mass are labeled as $x_{\rm com}$, $y_{\rm com}$, and $z_{\rm com}$. We also view some results in a rotated frame such that the instantaneous orbital separation vector defines the $x'_{\rm com}$-axis. In all cases, as described above, the unit length is the original donor-star radius, $R_1$.

For the current simulations we implement two small updates to the method described in \citet{2018ApJ...863....5M}. First, we adopt a different profile for the hydrostatic ``background'' material in the initial condition such that the background has several orders of magnitude lower total mass (outside the donor star) than in our previous models, approximately $6\times10^{-7} M_1$. We achieve this by adopting a lower background sound speed ($1/3$ rather than 1.0 in code units) such that the scale height is smaller. We join to the polytropic profile at a higher pressure ($10^{-6}$ rather than $10^{-8}$ in code units) such that the numeric scheme still is able to preserve the hydrostatic equilibrium of the stellar limb (compare to Figure 2 of \citet{2018ApJ...863....5M}). This change ensures that interaction with the ``background'' material of the initial condition does not play a role in the outflow properties presented here.  

Secondly, we modify an azimuthal zone-averaging scheme adopted by \citet{2018ApJ...863....5M} near the poles of the spherical-polar mesh. We average conserved quantities, in the $\phi$-direction, across the several zones nearest the poles, in order to avoid very distorted zone shapes (and associated courant-condition restrictions on the time step). In testing, we found that flow is more free to pass through the poles (reducing the presence of any numerical artifacts associated with the coordinate choice) if we average the vector, rather than scalar, momenta across zones. We therefore implement that approach here.  Despite these methodological improvements, we find little change in our qualitative results, confirming that the results presented here are not sensitive to either the initial condition or the treatment of the coordinate pole.

We will analyze the energetics of gas in the simulation domain extensively in what follows \citep[see][for a similar analysis of the energetics of particle-hydrodynamic simulations of binary interaction]{2016MNRAS.460.3992N}.
 In our simulated system, we evaluate the specific binding energy of material using the Bernoulli parameter,
\beq\label{bernoulli}
\mathcal{B} = \Phi + h + \varepsilon_{\rm k} 
\eeq
where $\Phi$ is the total potential from the binary,
 $\varepsilon_{\rm k}$ is the specific kinetic energy, and $h$ is the specific enthalpy.   
The total potential is 
\beq
\Phi = \Phi_1 + \Phi_2 + \Phi_{\rm sg},
\eeq
the combined gravitational potential of the donor and accretor core particles, respectively, and the self-gravitational potential of the gas. The potential of the accretor is that of a spline-softened point mass \citep[][equation A2]{1989ApJS...70..419H}. We treat the self-gravitational potential approximately as the potential of the undisturbed, spherical donor profile (as described in detail in \citet{2018ApJ...863....5M}, section 3.2.2).   The kinetic energy is 
\beq
\varepsilon_{\rm k} = {1\over 2}  v_{\rm com}^2,
\eeq
where $v_{\rm com}$ denotes the inertial-frame velocity relative to the system center of mass. The specific enthalpy is 
\beq
h = \frac{\gamma P}{(\gamma-1)\rho},
\eeq
where $P$ is the gas pressure and $\rho$ is the density.  We note that $\mathcal{B}$ relates to the specific total energy, $\varepsilon_{\rm tot}$, with $\varepsilon_{\rm tot} + P/\rho = \mathcal{B}$, where the addition of $P/\rho$ represents the potential energy associated with gas pressure's ability to do work along a free streamline.

\section{Mass Loss Preceding Stellar Coalescence}

Our simulation begins with Roche lobe overflow from a giant-star toward a less-massive, more-compact accretor. Material is lost from the donor  and streams into the accretor's vicinity. 
This mass loss from the donor is unstable because it runs away with ever-increasing rapidity and eventually leads to the engulfment of the accretor within the envelope of the donor. \citet{2018ApJ...863....5M} describe the dynamics of this runaway binary coalescence in detail.

Here we show that unstable mass transfer leads to overflow from the vicinity of the outer Lagrange points of the binary system, and trace the large-scale distributions that form. We examine the orbital-transformation mass-loss connection through the angular momentum carried away by material that is lost.

\subsection{Mass Transfer and Loss to the Circumbinary Enviornment}

\begin{figure*}[tbp]
\begin{center}
\includegraphics[width=0.99\textwidth]{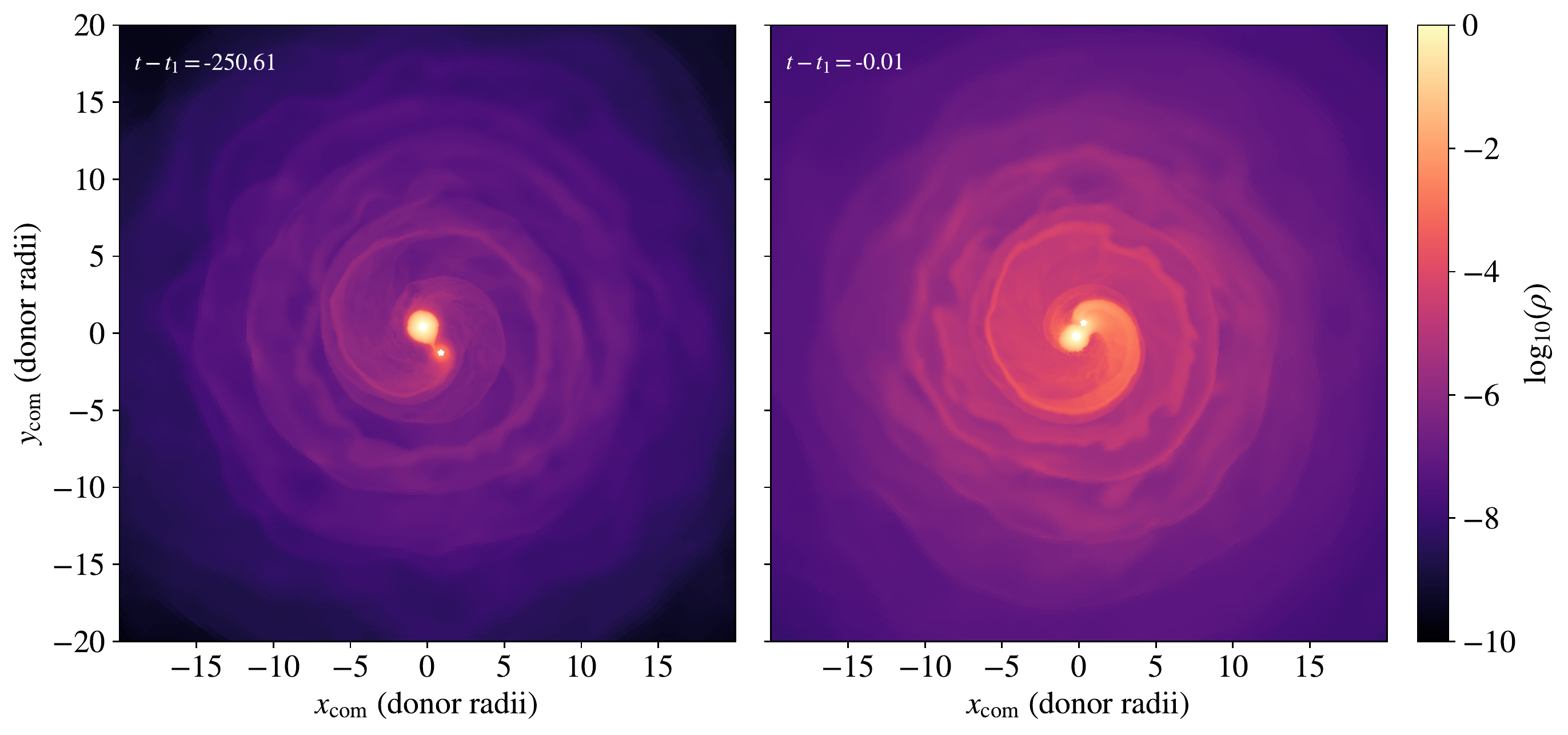}
\includegraphics[width=0.99\textwidth]{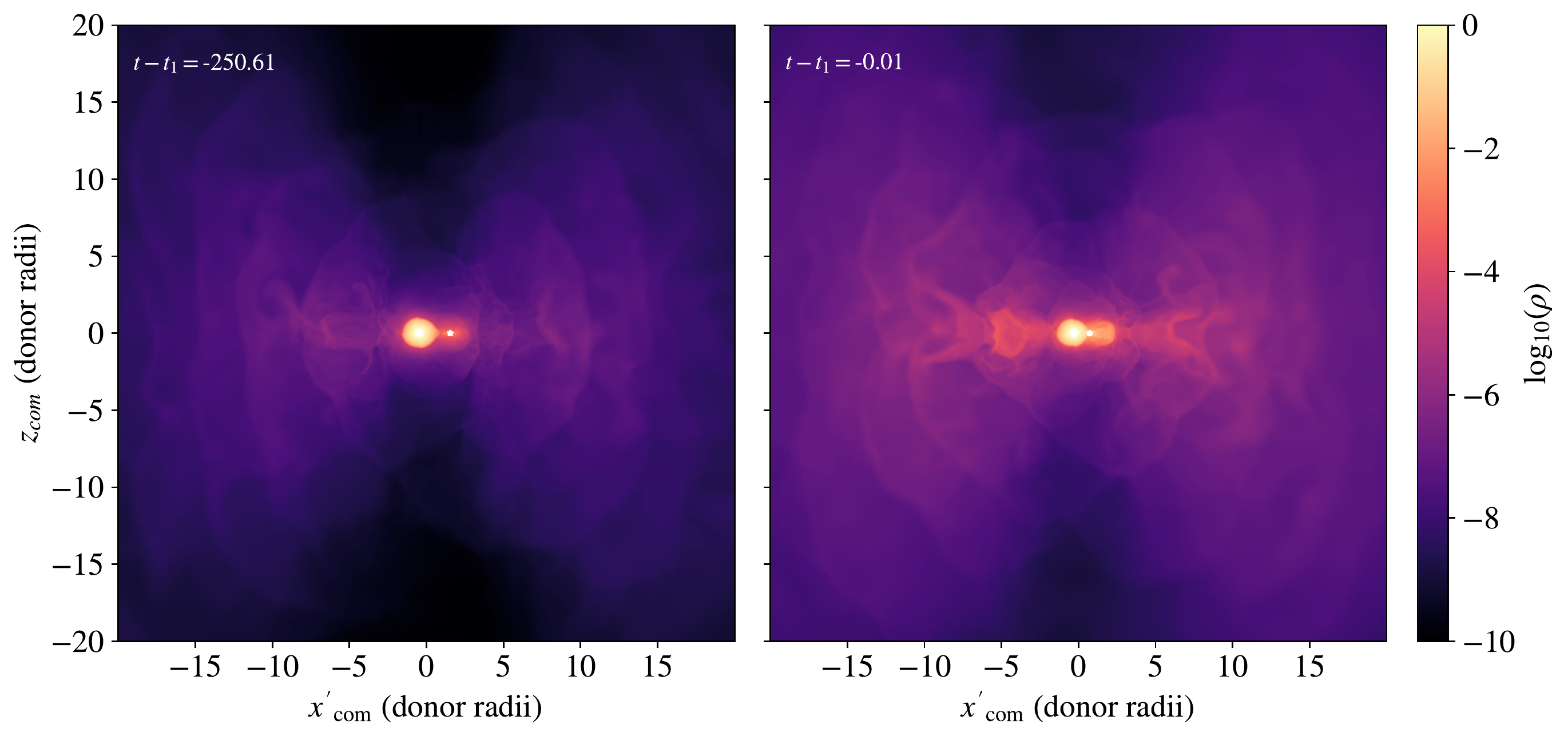}
\caption{Slices of gas density through the orbital midplane (upper panels) and perpendicular to the midplane (lower panels), extending $\pm20$ donor radii. As the two stars orbit toward their coalescence, continuous mass-shedding surrounds them in a toroidal cocoon of material, flung outward from -- and imparted energy and angular momentum at expense of -- the inner binary's orbit. Times are labeled in donor-star dynamical times, from several hundred donor dynamical times prior to merger (tens of orbits) to the moment when the accretor plunges within the donor's original radius. }
\label{fig:largescale}
\end{center}
\end{figure*}

In these simulations, we do not follow  the details of the accretion onto the compact star, in that we do not include cooling and we do not represent the accretor as a sink particle.  Instead, our simulations model the physical scenario of runaway mass transfer greatly exceeding the rate at which new mass may be incorporated into the accretor \citep[e.g.][]{2015MNRAS.449.4415P,2017MNRAS.465.2092P}. While some of the gas transferred from the donor orbits the accretor, much of the material streams from the vicinity of the binary and into the circumbinary environment, as shown in Figure \ref{fig:largescale} \citep[see also the numerical modeling of][]{1998MNRAS.300...39B,2005Ap&SS.296..391B,2007ARep...51..836S}. The panels of Figure \ref{fig:largescale} show the circumbinary gas distribution at two times, from several hundred donor dynamical times before, to the moment of, the engulfment of the accretor within the donor's envelope \citep[and may be compared to recently published particle-hydrodynamic simulations by][see their Figures 2, 8]{2018arXiv180902297R}.  The upper panels slice through the orbital midplane, the lower panels perpendicular to the midplane.

Material is lost from the vicinity of the binary preferentially in the orbital midplane \citep{1998MNRAS.300...39B}. The location of preferential mass loss from the binary is near the second Lagrange point, $L_2$, which has the lower potential of the outer Lagrange points.  In the midplane, spiral features of mass loss form because material's angular velocity slows as it expands, imprinting the orbital motion of the binary on the outflow \citep[e.g.][]{1979ApJ...229..223S,1998MNRAS.300...39B,2009ARep...53..223S}.  Because new outflows are constantly forming, there is significant self-interaction among the outflowing gas \citep{1998MNRAS.300...39B,2009ARep...53..223S,2007ARep...51..836S,2016MNRAS.455.4351P,2016MNRAS.461.2527P,2017ApJ...850...59P,2017MNRAS.471.3200M}. The circumbinary structure is continuously subject to new spiral shocks racing through from new outflow. Dense boundary layers can be seen within the streams that separate unshocked ejecta, from shocked ejecta, from the leading edge of shocks racing into new material. Within this bifurcated structure instabilities cause streams to buckle, fragment, and billow, racing outward, stalling, then falling back in \citep[as seen also in the simulations of][]{2009ARep...53..223S}. 

Perpendicular to the orbital plane, we observe that the circumbinary gas forms a toroidal distribution that also bears the signatures of gas self-interaction. While mass loss is strongly equatorially concentrated, self-interactions between outflow components widen the opening angle of outflow material surrounding the binary. Rather than remaining equatorially concentrated, material extends into the thick, circumbinary torus seen in the lower panels of Figure \ref{fig:largescale}.  Only the polar regions remain relatively evacuated, with the expelled gas having too much angular momentum to significantly populate these areas.

\subsection{Orbital Decay and Cumulative Mass Loss}

\begin{figure}[tbp]
\begin{center}
\includegraphics[width=0.48\textwidth]{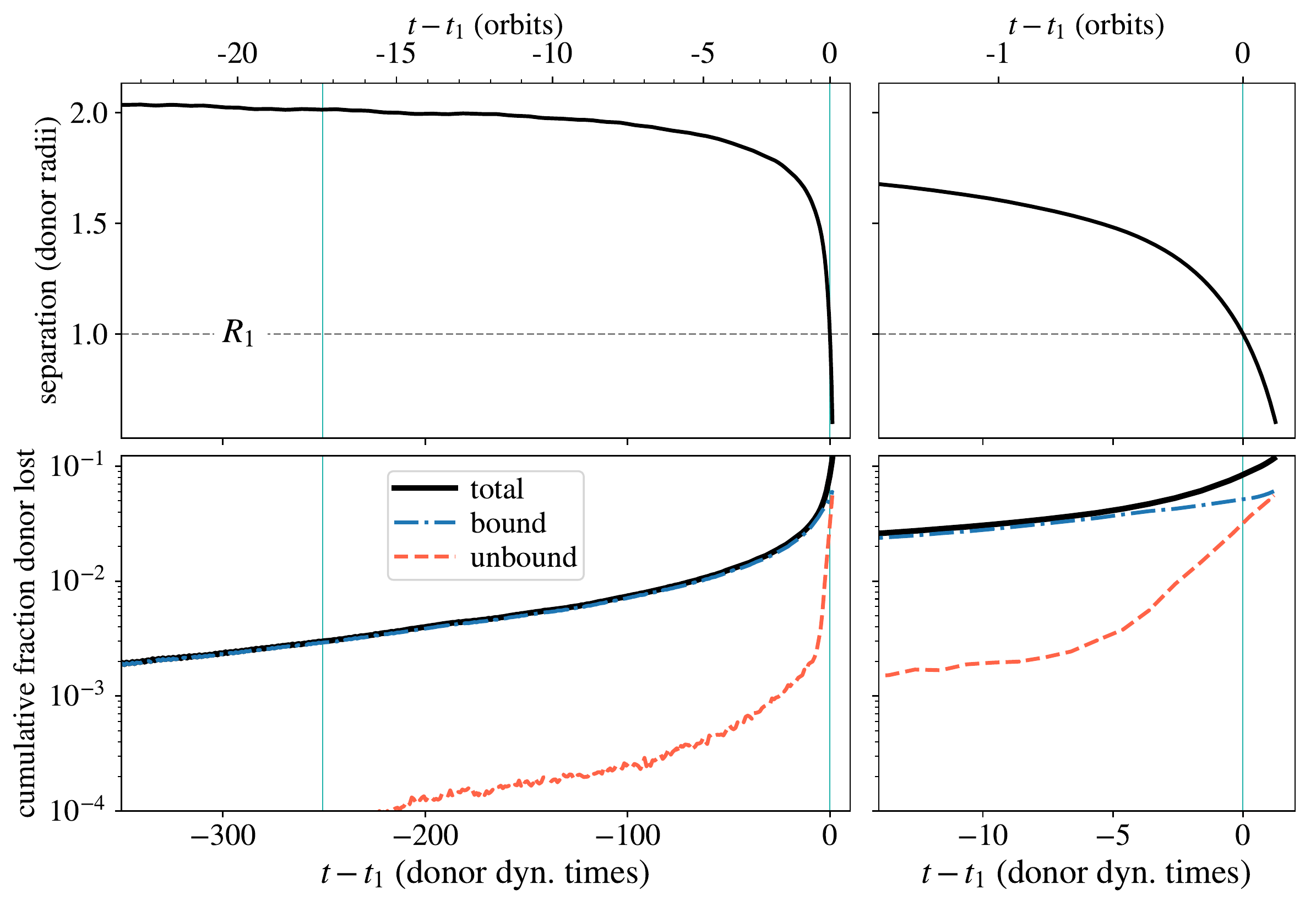}
\caption{ Binary separation (upper panels) and cumulative mass loss from the donor (lower panels) as a function of time.  Vertical lines mark the times of the snapshots shown in Figure \ref{fig:largescale}. The total mass lost from the donor increases as the binary spirals closer. The mass loss rate and orbital decay rate both grow significantly leading up to the plunge of the accretor within the envelope of the donor. 
We show the decomposition of cumulative mass loss into bound and unbound components. Bound mass loss dominates the early inspiral, while significant quantities of unbound mass are generated only in the final several dynamical times prior to coalescence. 
}
\label{fig:masslosstime}
\end{center}
\end{figure}

As the binary traverses from the Roche limit separation toward coalescence, mass is continuously lost from the donor star. The exchange of angular momentum between the binary orbit and the donor mass that is extracted, mediated by gravitational torques, defines this phase of orbital decay \citep{1956AJ.....61...49H,1963ApJ...138..471H,1963ApJ...138..481H}. Here we elaborate on the connection between orbital transformation and cumulative mass lost to the circumbinary environment. 

Figure \ref{fig:masslosstime} shows the binary separation as a function of time in the upper panels and the cumulative mass removed from the donor star in the lower panels. The entire duration of the model calculation is approximately 500 donor dynamical times or 34 initial orbital periods. The right-hand panels zoom in on the final few dynamical times as the separation, measured between the accretor and the donor core, plunges from exterior to the initial donor radius to interior. Orbital decay begins slowly but accelerates as the binary separation tightens. Similarly, mass loss from the donor first accumulates slowly, but then becomes extremely rapid in the final orbits, as the donor star increasingly overflows its Roche lobe.

\begin{figure}[tbp]
\begin{center}
\includegraphics[width=0.48\textwidth]{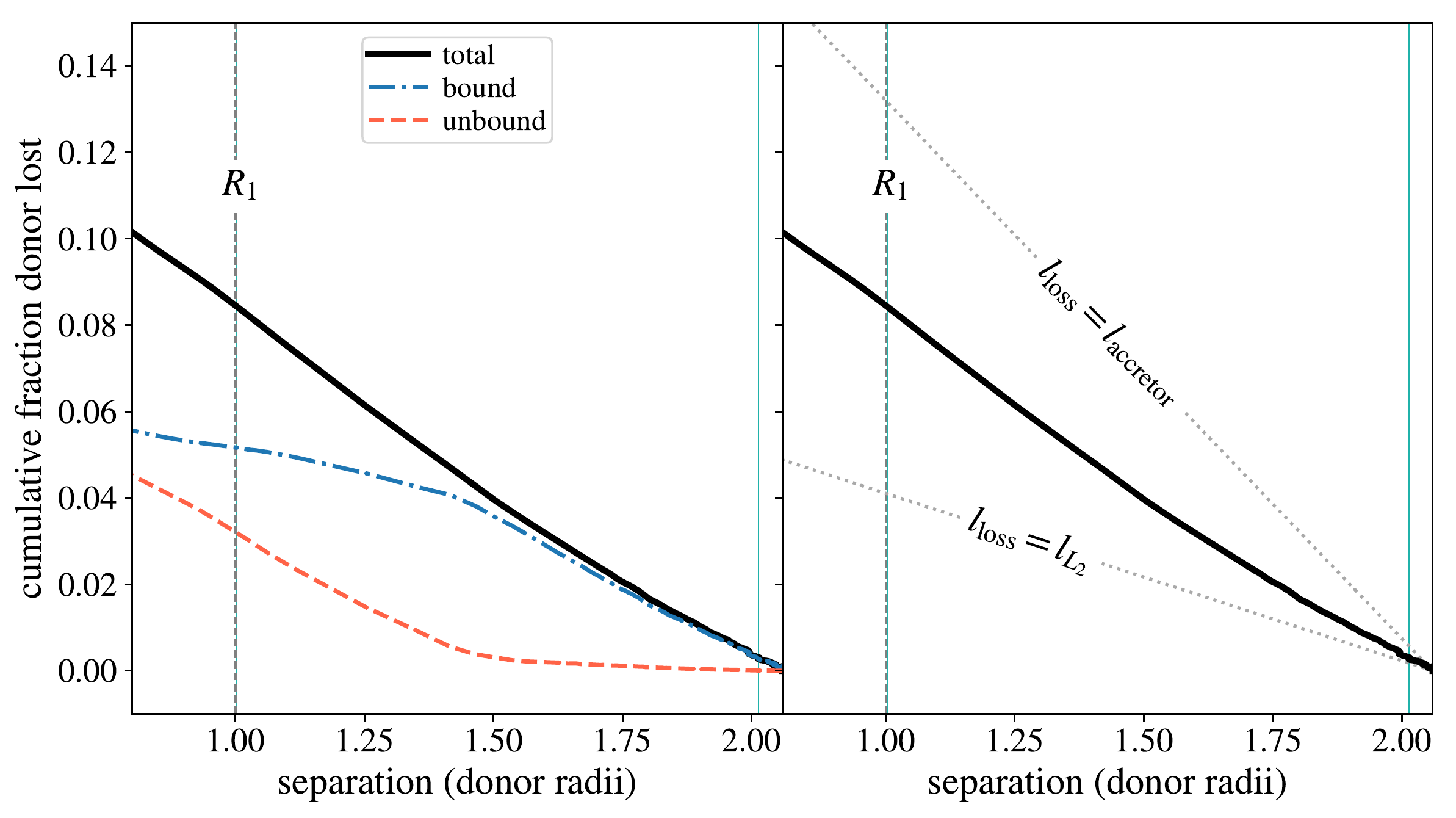}
\caption{ Cumulative mass lost from the donor star (as shown in the lower panels of Figure \ref{fig:masslosstime}) as a function of orbital separation. The left panel shows the total mass lost decomposed into material that remains bound to the binary and material that is unbound. The right panel notes the connection between mass loss and orbital transformation by comparing to two representative values of the the specific angular momentum carried by outflowing material -- that of the $L_2$ point and that of the accretor. Material is lost with average specific angular momentum intermediate between these two values. The total mass lost from Roche lobe overflow to the engulfment of the accretor within the envelope of the donor is roughly 8\% of the mass of the donor (approximately 25\% that of the accretor), of which approximately two-thirds is bound and one-third is unbound.  }
\label{fig:masslosssep}
\end{center}
\end{figure}

\citet{2018ApJ...863....5M} showed that a semi-analytic model of mass loss from the donor star and orbital evolution reproduces the key features of the simulated coupled mass exchange and orbital evolution. Similarly, in a series of works, \citet{2014ApJ...788...22P,2017ApJ...850...59P} have shown that such a model explains the observed time-evolving orbital decay of the binary V1309 Sco leading up to its coalescence. \citet{2017ApJ...850...59P} also performed radiative transfer modeling that shows that an outflow from $L_2$ can explain the main features of the OGLE eclipsing lightcurve of V1309 Sco, which evolved significantly over the years prior to the outburst \citep{2011A&A...528A.114T}. 

We can understand the key features of these models, as they pertain to mass loss to the circumbinary environment as follows. In being removed from the donor star and expelled from the binary, material acquires a certain specific angular momentum, which we denote $l_{\rm loss}$, that is larger than its original specific angular momentum at $L_1$ \citep{1963ApJ...138..471H}. This gain in specific angular momentum is mediated by gravitational torques, primarily from the accretor star, and comes at the expense of the orbital angular momentum \citep{2017ApJ...835..282M,2018ApJ...863....5M}. 

As a result, a given amount of mass loss drives a particular change in orbital angular momentum,   
\beq\label{dmdonor}
\Delta M_{\rm donor} \approx  \frac{\Delta L_{\rm orb} }{l_{\rm loss}}.
\eeq
For example, the transition from the Roche limit separation ($2.06R_1$ in our example where $q=0.3$) to engulfment of the accretor, $a\approx R_1$, implies a change in orbital angular momentum, 
\beq
\Delta L_{\rm orb} =   L_{\rm orb}(R_1) - L_{\rm orb}(a_{\rm RL}).
\eeq 
Note that because $\Delta L_{\rm orb}<0$, the change in donor mass is also negative -- the donor-star loses mass to the surroundings via the outflow \citep[see a similar model applied to the binary V1309 Sco in][]{2014ApJ...788...22P}. 

If $l_{\rm loss}$ is constant, and additionally if the binary has relatively-constant mass, this implies a relationship between orbital separation and total mass lost. This relationship is shown in Figure \ref{fig:masslosssep}. Without a means to predictively estimate the specific angular momentum of the outflow, the chance to measure this quantity using numerical simulations is one of the key results from the modeling in \citet{2018ApJ...863....5M}. The slope of mass loss versus separation is related to the specific angular momentum of the outflow, as described in equation \eqref{dmdonor}. For guidance, we show the mass loss that would result from two characteristic specific angular momenta of the binary, that of the accretor star and that of the $L_2$ Lagrange point \citep{1998CoSka..28..101P}. The $L_2$ Lagrange point is a larger lever arm, and thus represents a higher specific angular momentum. As a result, the predicted mass loss is lower than that with the accretor's specific angular momentum.  

Our simulated result shows that as the binary separation changes from the Roche limit to engulfment, the donor loses approximately 8\% of its mass (or about 25\% of the accretor's mass). Assuming that material is lost with the specific angular momentum of the $L_2$ point, \citet{2017ApJ...835..282M}, followed a similar argument to estimate that binaries lose 10-15\% of the accretor's mass in traversing from the Roche limit to engulfment.  The simulated result therefore represents a specific angular momentum for the outflow that is lower than that of the $L_2$ outer Lagrange point, but higher than that of the accretor -- with higher resulting cumulative mass loss.  

Finally, Figures  \ref{fig:masslosstime} and \ref{fig:masslosssep} highlight a key result that we explore further in Section \ref{sec:boundunbound}, the breakdown of the total mass lost into bound ($\mathcal{B}<0$) and unbound ($\mathcal{B}>0$) components. We observe that the bulk of the mass lost while the binary is separated by $>1.5R_1$ is  bound to the binary, despite having been unbound from the donor star. As the separation shrinks, we observe a marked increase in unbound ejecta, such that by the time the separation is approximately $R_1$, roughly two-thirds of the cumulative outflow have been bound to the binary, while one-third is unbound.

\section{From Bound Outflows to Unbound Ejecta}\label{sec:boundunbound}

\begin{figure*}[tbp]
\begin{center}
\includegraphics[width=0.9\textwidth]{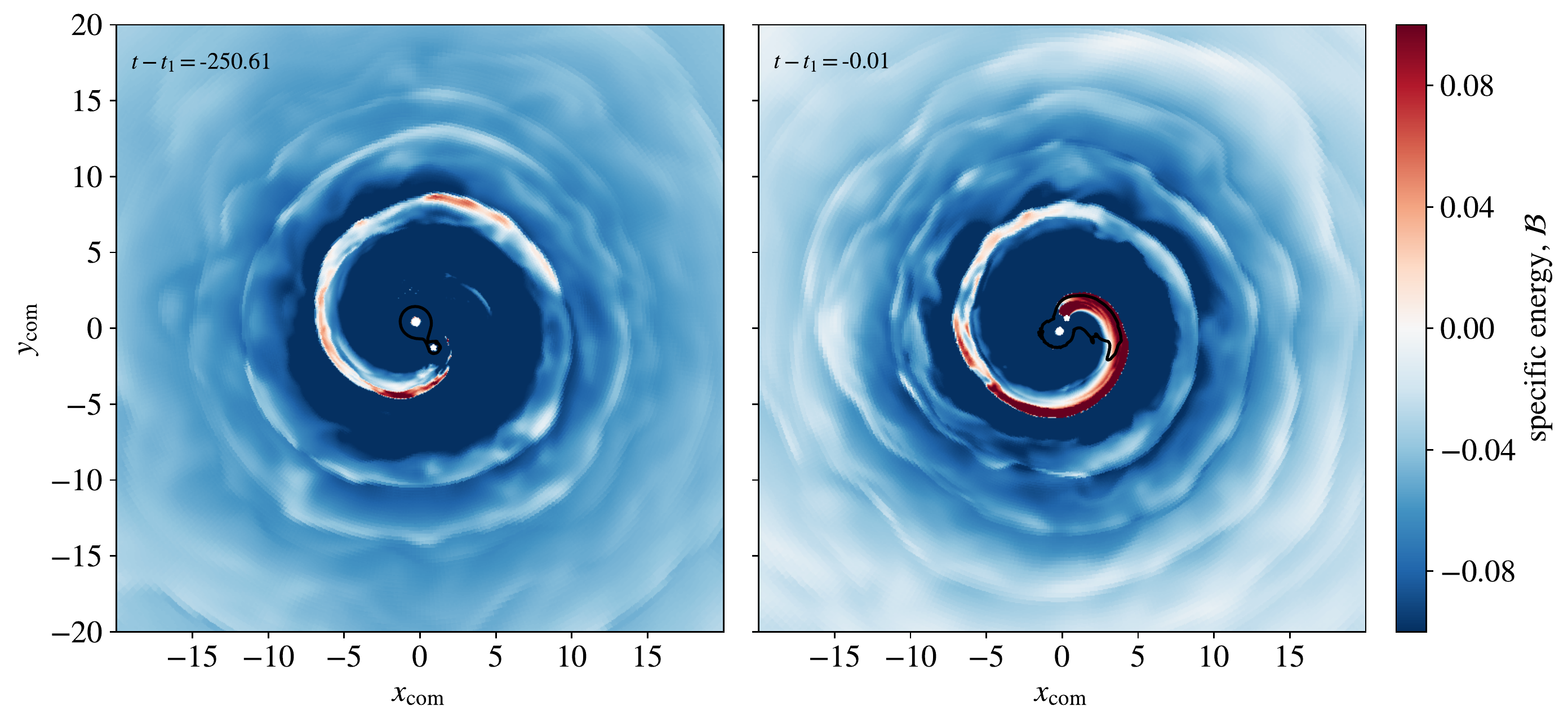}
\caption{Gas Bernoulli parameter -- the sum of specific kinetic energy, potential, and specific enthalpy -- in a slices through the orbital midplane corresponding to the same snapshots as Figure \ref{fig:largescale}. A single contour shows a density of $\rho=10^{-3}$. Nearly all of the material which has been carried away from the donor remains bound to the binary at early times ($\mathcal{B}<0$, left panel). As the binary coalesces, as shown in the later snapshot, a stream of unbound material is ejected ($\mathcal{B}>0$, right panel).  }
\label{fig:largescaleE}
\end{center}
\end{figure*}

Here we examine the transition in flow morphology within the binary: initially the interaction leads to the creation of a bound outflow that settles into a circumbinary torus; at later stages the interaction leads to the generation of unbound ejecta. In Figures \ref{fig:masslosstime} and \ref{fig:masslosssep}, we have shown the connection of this transition to orbital decay of the binary. In Figure \ref{fig:largescaleE}, we slice through the orbital midplane and plot gas  specific energy, $\mathcal{B}$, for the same snapshots as shown in Figure \ref{fig:largescale}. Negative values represent material bound to the binary, while positive values represent unbound material. The majority of material in the circumbinary environment remains bound to the binary until an unbound stream of is generated late in the binary's coalescence, as shown in the second snapshot.

\subsection{Predictions from Test-Particle Trajectories}\label{sec:testparticle}

\begin{figure}[tbp]
\begin{center}
\includegraphics[width=0.45\textwidth]{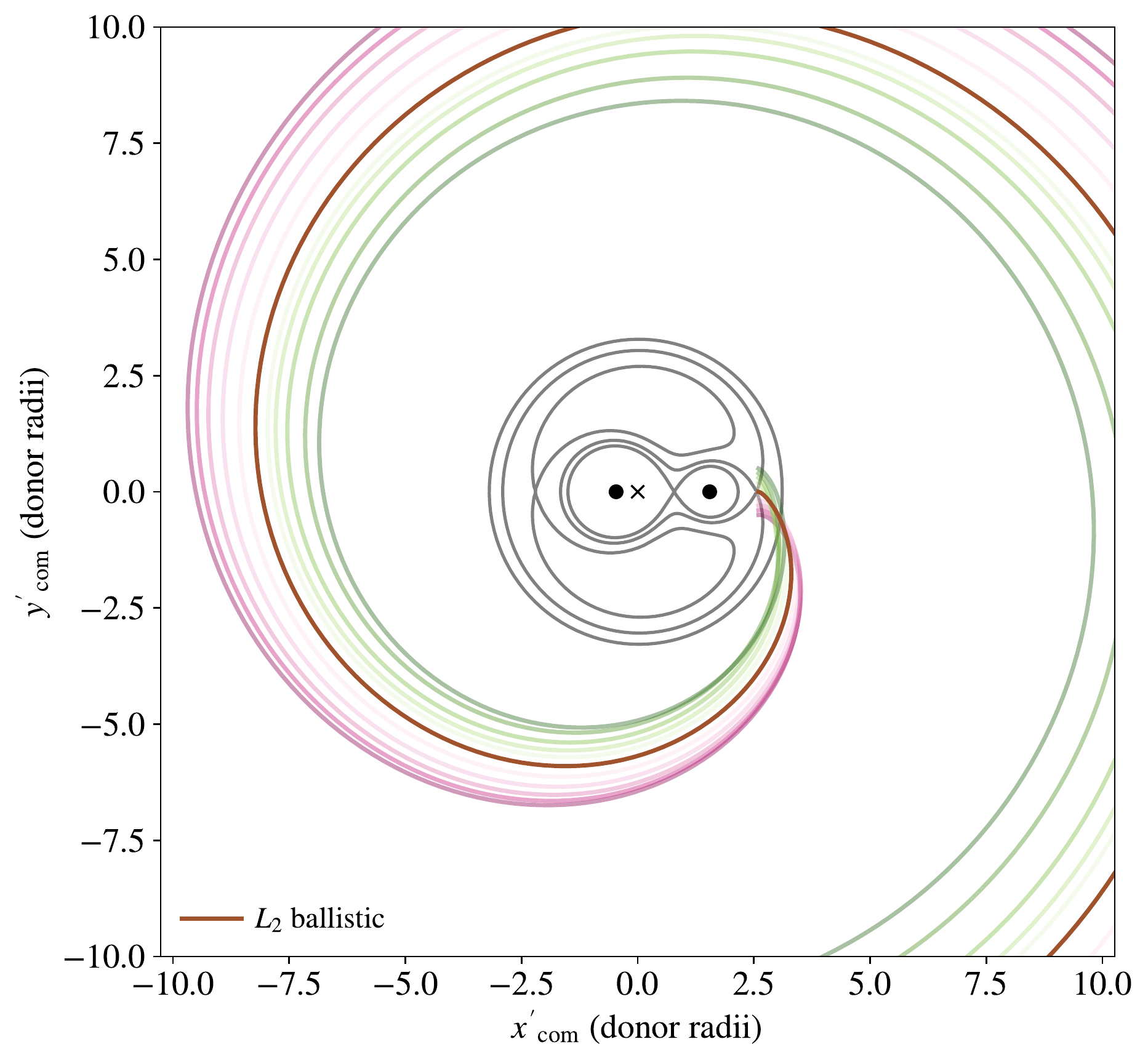}
\includegraphics[width=0.49\textwidth]{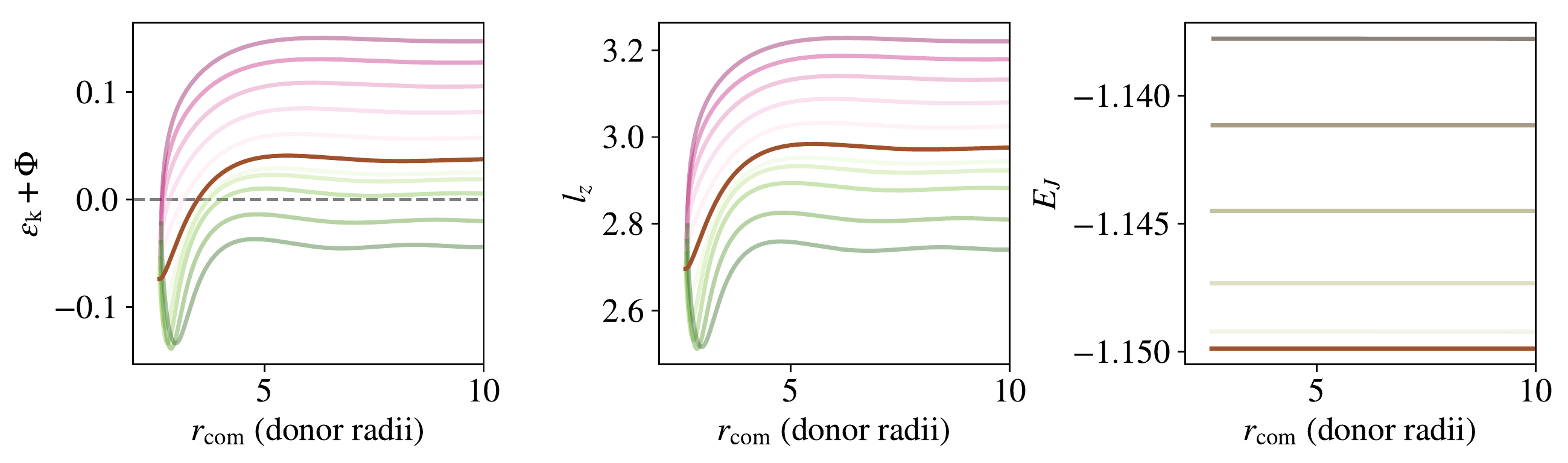}
\caption{ Test-particle trajectories in the binary effective potential.  The upper panel shows the trajectory in the binary orbital plane, and the lower panel the test-particle specific energy, $\varepsilon_{\rm k} + \Phi$, angular momentum, and Jacobi parameter of particles initialized in corotation just outside the $L_2$ point of the binary (brown line). Colored lines share the same $x'_{\rm com}$ coordinate but have a range of $y'_{\rm com}$-axis offsets  leading or trailing $L_2$. The trajectory from $L_2$ is accelerated to positive energy after approximately a half-orbit. Those originating on the trailing edge have larger final energies, those on the leading edge are decelerated initially and retain negative total energy.  }
\label{fig:testparticle}
\end{center}
\end{figure}

We begin by discussing test-particle trajectories in the binary potential, because these serve as a useful context for our investigation of the bound and unbound gas-dynamical flows generated as the binary coalesces. 

\citet{1979ApJ...229..223S}, and later \citet{2016MNRAS.455.4351P,2016MNRAS.461.2527P,2017ApJ...850...59P} have examined the idealized problem of mass loss from the outer, $L_2$ Lagrange point in a binary. The $L_2$ point is the lower of the two outer saddle points of the effective gravitational potential in the corotating frame of the binary -- therefore, material in corotation there is free to slide away from the binary. In Figure \ref{fig:testparticle}, we recreate similar test-particle trajectories by integrating the motion of particles in the effective potential of the binary, taking the same separation and masses as shown in the left panels of Figure \ref{fig:largescale}, roughly 2 donor radii. We highlight the trajectory of a particle initialized just outside $L_2$ in brown, and those initialized offset by a range of positions in $y'_{\rm com}$ spanning $\pm 0.5R_1$ in a color spectrum. 

The basic finding of \citet{1979ApJ...229..223S} is that test particles emitted in corotation at $L_2$ become unbound for most binary mass ratios, $0.064<q<0.78$ (see their Section V).   From Figure \ref{fig:testparticle}, we see that the trajectory initialized just outside $L_2$ (thick brown line) is not unbound initially, instead it is accelerated by the forces from the two stars until it is unbound eventually. By several times the binary separation, approximately about $5R_1$, or a half-orbit in the corotating frame, the particle reaches close to its asymptotic energy.  Over this same scale, the specific angular momentum, $l_z$, increases to its asymptotic value. 

For ballistic trajectories in a time-invariant but rotating binary potential, energy and angular momentum are related by the Jacobi parameter,
\beq\label{jacobi1}
E_J = {1 \over 2} |{\bf \dot r_{\rm rot}}|^2  + \Phi - {1\over 2} | \Omega \times {\bf r_{\rm rot}} |^2,
\eeq
expressed here in terms of the separation vector to the system center of mass in the corotating frame, $\bf r_{\rm rot}$  \citep{2008gady.book.....B}.
The Jacobi parameter is conserved along ballistic trajectories. It may also be re-written in several other, informative forms, 
\begin{align}\label{jacobi2}
E_J & = \varepsilon_{\rm k} + \Phi - \Omega l_{z}, \nonumber \\
       & = {1 \over 2} |{\bf \dot r}|^2 +  \Phi_{\rm eff},
\end{align}
where $\varepsilon_{\rm k} + \Phi$ is the inertial-frame specific energy for a test particle (which has no thermal energy), $l_z$ is the $z$-component of the inertial-frame specific angular momentum, and finally $\Phi_{\rm eff} = \Phi - {1\over 2}\Omega^2 R^2$ is the effective potential of the rotating frame. 
The first equality of equation \eqref{jacobi2} shows that changes in energy and angular momentum occur in tandem such that $E_J$ is conserved. The second equality shows that $E_J$ may be written in a form similar to a total energy (kinetic plus effective potential) in the corotating frame.  The lower right panel of Figure \ref{fig:testparticle} shows $E_J$ along the trajectories plotted. Its value is constant along a given trajectory.

Figure \ref{fig:testparticle} also shows the trajectories of particles initialized in corotating at the same $x'_{\rm com}$ coordinate as $L_2$ but offset in $y'_{\rm com}$. The leading-edge particles ($y'_{\rm com}>0$) are first decelerated as they pass by the accretor then are never accelerated to positive total energy. On the other hand, the  particles initialized on the trailing edge  ($y'_{\rm com}<0$) eventually reach more positive energy than that initialized at  $L_2$. Similar variations in the asymptotic specific angular momentum occur. The lower right panel of Figure \ref{fig:testparticle} shows that different trajectories have different Jacobi parameter, but that it varies symmetrically between trajectories with a given $\pm y'_{\rm com}$ offset.  Finally, we note that the upper panel of Figure \ref{fig:testparticle} shows that the relative deceleration and acceleration of leading and trailing particles, respectively, implies that the trajectories cross just outside of $L_2$. 

\citet{2016MNRAS.455.4351P,2016MNRAS.461.2527P} extended these considerations of test particles to studies of the gas dynamics of outflows initialized with low sound speed from a nozzle at $L_2$. They study the thermodynamics of these flows in extensive detail. With respect to the kinematics, they find trajectories that closely mimic those of test particles because they inject particles into their simulation from a narrow nozzle near $L_2$ (i.e. very close to the brown curve in Figure \ref{fig:testparticle}, which starts bound and becomes unbound).  Under these assumptions, \citet{2016MNRAS.455.4351P} report largely-unbound gas-dynamical outflows for the same range of mass ratios, $0.064<q<0.78$, estimated by \citep{1979ApJ...229..223S}, see \citet{2016MNRAS.455.4351P}'s Figure 3. The largest predicted asymptotic velocities occur for $q\approx 0.3$, the same value adopted in our simulations \citep[][Figure 3]{1979ApJ...229..223S,2016MNRAS.455.4351P}. Figure \ref{fig:testparticle} shows, however, that even gas ejected from relatively near $L_2$ (with positive $y'_{\rm com}$) never becomes unbound. 

\subsection{Bound Outflow Phase}\label{sec:boundoutflow}

An outflow, unbound from either component but bound to the binary, is generated during early coalescence of the binary, at stages following Roche lobe overflow, as shown in the midplane slice of Figure \ref{fig:largescaleE}. This bound mass loss is the dominant component until the final few dynamical times prior to the engulfment of the accretor within the envelope of the donor (Figure \ref{fig:masslosstime}) or, equivalently, at separations larger than roughly $1.5R_1$ (Figure \ref{fig:masslosssep}) in our simulated system. 

\subsubsection{Flow Morphology}

\begin{figure*}[tbp]
\begin{center}
\includegraphics[width=0.38\textwidth]{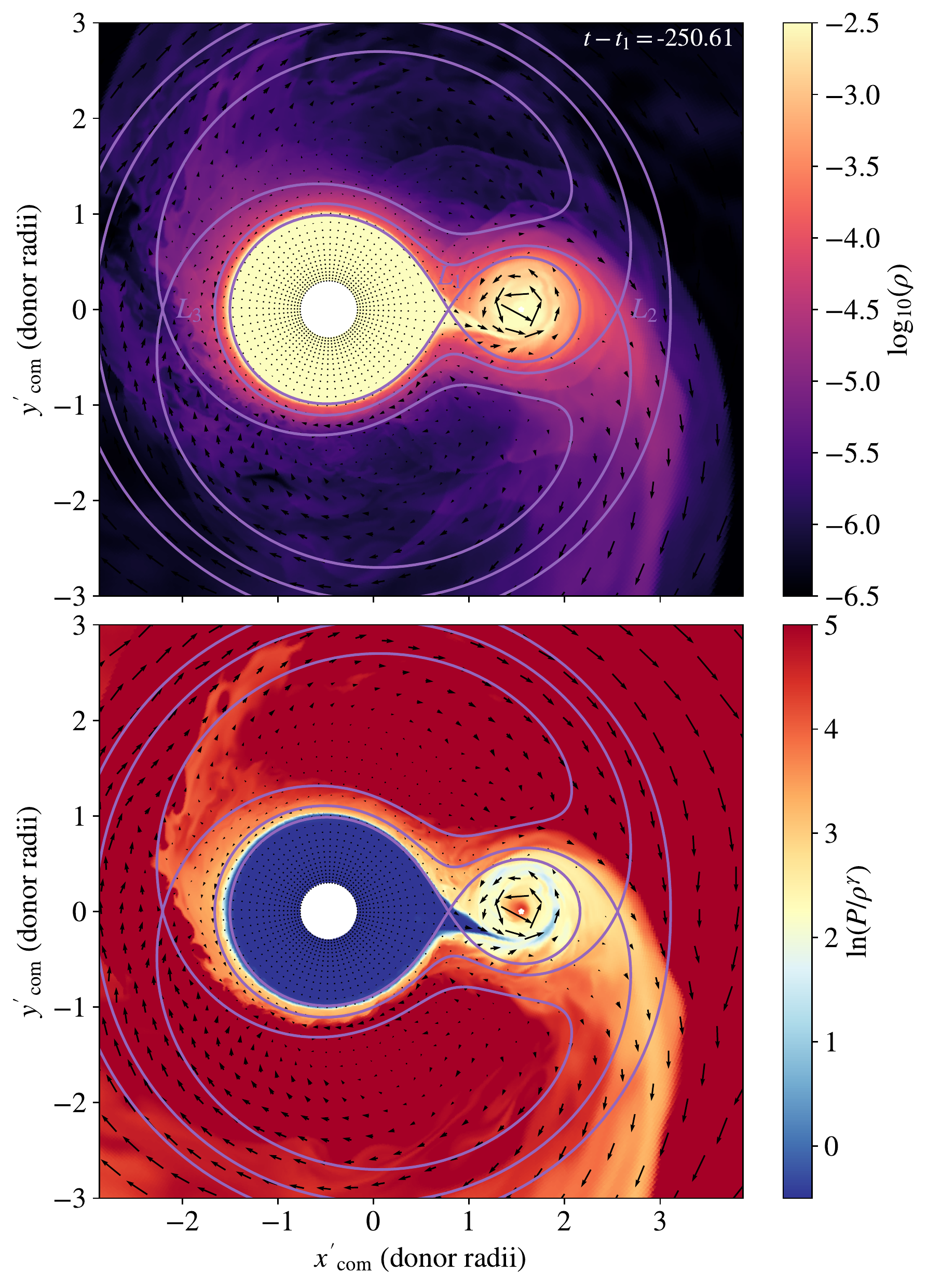}
\hspace{1cm}
\includegraphics[width=0.55\textwidth]{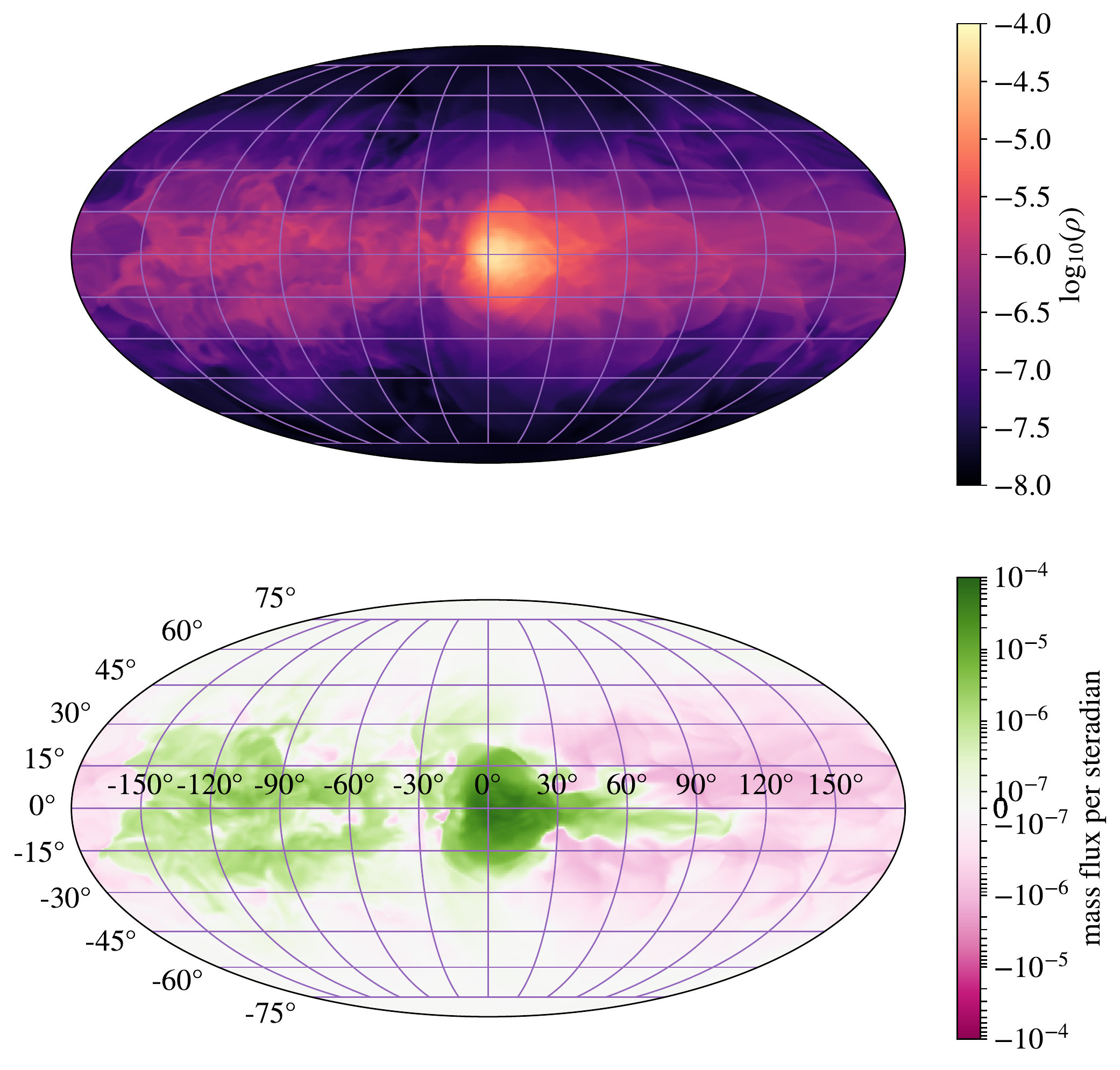}
\caption{Slices through the orbital midplane, taken from the early-time snapshot, rotated such that the orbital separation vector is along the $x$-axis (left) and slices at constant radius from the donor (right).  {\it Left panels:} The upper-left panel shows gas density, the lower-left specific entropy.  Material is pulled from the donor star as it overflows its Roche lobe into the Roche lobe of the accretor. Gas forms a rotating disk about the accretor. Rather than piling up or accreting, the majority of this gas is then expelled from the vicinity of the accretor in a broad stream near the $L_2$ outer Lagrange point.  {\it Right panels:} Slices of constant radius relative to the donor star, selected to intersect $L_2$. Zero degrees in altitude corresponds to the orbital midplane, while zero degrees in azimuth is the location of $L_2$.  The trailing direction (relative to binary motion) corresponds to positive azimuthal coordinate. The upper-right panel shows density along this surface, the lower-right panel mass flux across the surface. Material is concentrated near the location of $L_2$, as is the majority of outflow -- primarily in a tens of degrees locus. Material is present, however, at all azimuthal angles (inflowing or outflowing) and within roughly 30 degrees of the equator.  
}
\label{fig:binaryflow0}
\end{center}
\end{figure*}

Following Roche lobe overflow,  material is pulled from the donor star. The left panels of Figure \ref{fig:binaryflow0} show density and specific entropy in the vicinity of the mass-transferring binary in a slice through the orbital plane.  This is the earlier of two snapshots previously shown, 250 donor dynamical times prior to merger, when the binary separation is 2.01 donor radii, or approximately 98\% percent of the Roche limit separation (the time and separation are marked in Figures \ref{fig:masslosstime} and \ref{fig:masslosssep}). We additionally plot gas velocity relative to the instantaneously corotating frame (vectors) and equipotential contours that intersect the $L_1$, $L_2$, and $L_3$ Lagrange points. 

Transferred material streams into an accretion disk about the accretor via a dense, narrow, low-entropy flow \citep{1975ApJ...198..383L}. It collides with gas already orbiting the accretor and shocks, increasing the specific entropy of the gas in this rotating flow \citep[e.g.][]{1998MNRAS.300...39B}. As the process continues, pressure gradients force gas outward in all directions away from the accretor, finding approximate balance with gravity. The highest-entropy gas is pushed to the surface, where it flows away from the accretor -- either back toward the donor, or away from the binary \citep{1998MNRAS.300...39B,2007ARep...51..836S,2009ARep...53..223S}. 

Relatively rapidly, the entire surroundings of the binary are filled with material drained from the donor. The donor and the disk about the secondary orbit within this circumbinary material.  Higher-entropy surface layers surround both stars, roughly following the equipotential surfaces of the Roche lobes \citep{1979ApJ...229..223S}. These surface layers are particularly compressed on the leading edges of the two binary components, which continuously plow into new material. Within these surface layers, gas slides around the stars and forms tails that unwrap in spirals from the vicinity of the two outer Lagrange points, $L_3$, near the donor, and, especially, $L_2$, near the accretor.

In the right-hand panels of Figure \ref{fig:binaryflow0}, we examine outflow from the binary more closely. These panels  show maps of gas density and mass flux across a spherical surface centered on the donor. The radius of the sphere, 3.01 donor radii, intersects the $L_2$ point. The orientation is as viewed from the donor center,  such that $L_2$ is at the origin of altitude and azimuthal angles; positive azimuthal angles correspond to the trailing direction of the outflow. The highest density material lies in a band within roughly $\pm$30 degrees of the equator. A strong concentration is present near $L_2$, with the maximum densities about two orders of magnitude higher than elsewhere along the equator. 

The lower-right panel of Figure \ref{fig:binaryflow0} shows the mass flux per unit steradian (in units of donor masses per donor dynamical time). Green represents outflow while pink represents inflow. What we see from this slice is that material is not uniformly outflowing. There are regions of inflow as well as regions of outflow that reflect the density distribution in their extent above and below the equatorial plane. Also mirroring the gas density, the most intense region of outflow lies in a broad locus near $L_2$.

\subsubsection{A Bound Outflow}

Having noted how flow within the binary gives rise to an outflow, here we analyze its properties in more detail.

\begin{figure}[tbp]
\begin{center}
\includegraphics[width=0.44\textwidth]{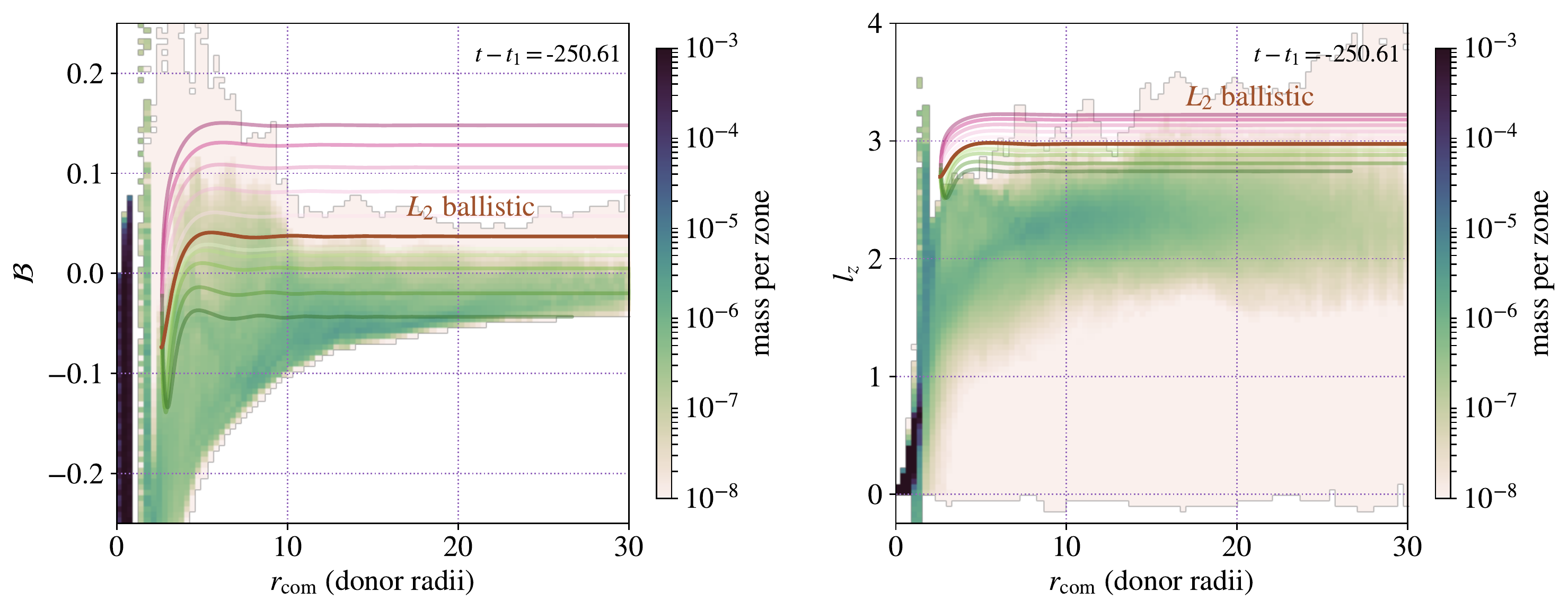}
\includegraphics[width=0.45\textwidth]{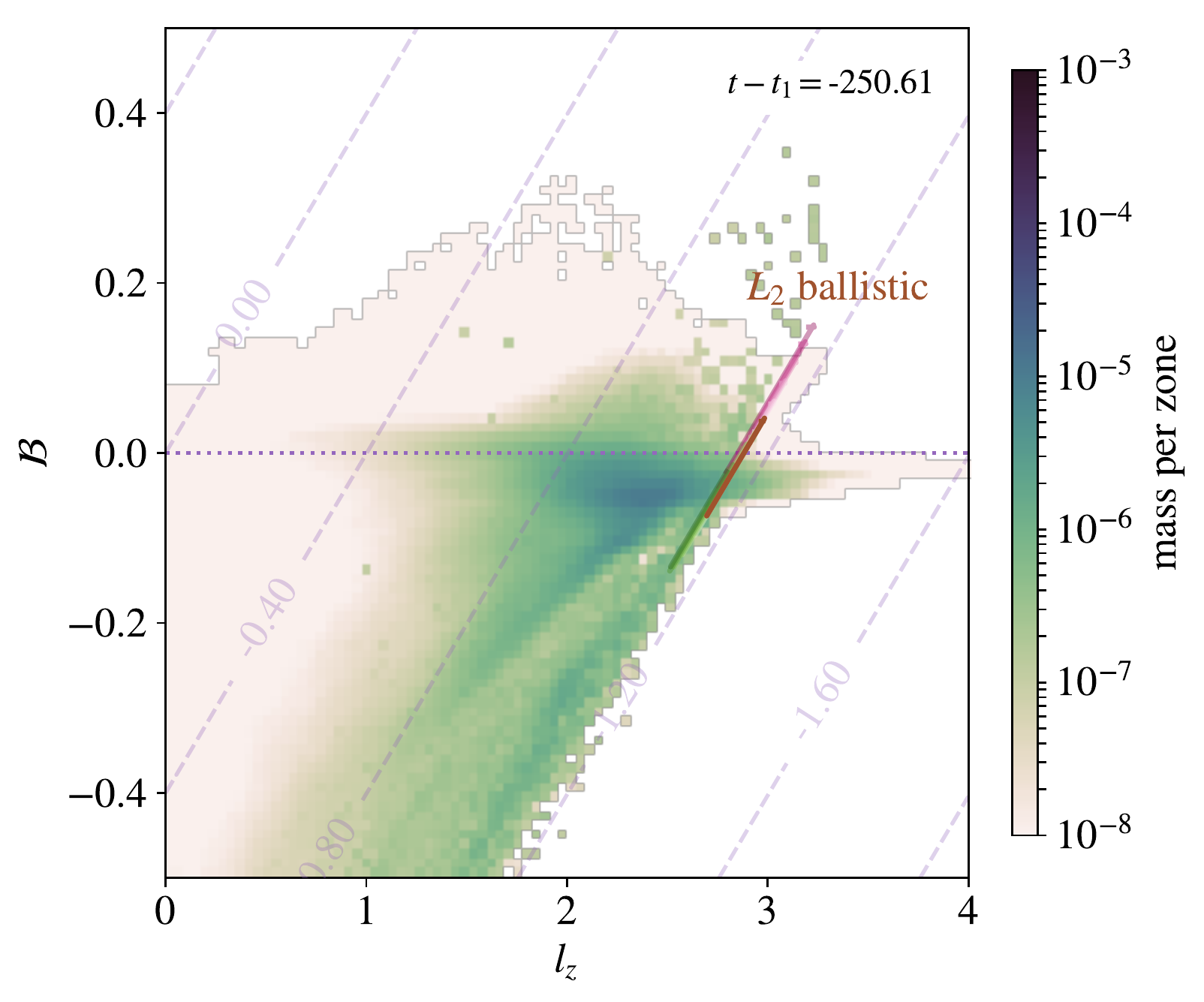}
\caption{Distributions of gas specific energy and angular momentum at early time. The upper panels show quantities as a function of radius from the system center of mass, the lower shows material outside of the $L_2$ point in energy-angular momentum phase space, with contours of constant Jacobi parameter shown in the background, which are valid in the limit of vanishing specific enthalpy (further comment in the text).   Each panel compares to the range of ballistic trajectories initialized from corotation near $L_2$ shown in Figure \ref{fig:testparticle}.  The upper panels show that initially-broad distributions of energy and angular momentum asymptote to relatively constant values at larger radii. Typical angular momenta are lower than those of any of the corotation trajectories. The typical phase space populated by the outflow (lower panel) is mildly-negative energy and larger Jacobi parameter than any of the $L_2$ corotation trajectories. 
 }
\label{fig:bounddist}
\end{center}
\end{figure}

A closer inspection of Figure \ref{fig:largescaleE} reveals that outside of about 10 stellar radii, the specific binding energy of material is approximately constant, with negative values on the order of $\mathcal{B}\approx-0.05$ (where the characteristic unit for the specific energy is that of the donor: $GM_1/R_1$). The corresponding semi-major axis, relative to the binary center of mass, of this specific energy is $a\approx G(M_1+M_2) / 2 \mathcal{B} \approx 13R_1$. Thus the relatively weak binding energy of the circumbinary outflow corresponds to the fact that it expands to form a torus with typical size of tens of donor radii about the binary. 

We examine the properties of the bound outflow more quantitatively in Figure \ref{fig:bounddist}.  The upper panels show the distributions of specific energy quantified by the Bernoulli parameter (left) and specific angular momentum (right) of material in the entire simulated domain. Colors denote mass per zone within the histogram. For reference, we plot the ballistic trajectories initialized in corotation near $L_2$ (as in Figure \ref{fig:testparticle}). At small radii, material in the simulation domain has a broad range of energies and angular momenta. As it expands, however, it settles into  a narrowing range of energies and angular momenta that become relatively constant beyond approximately ten donor radii. Much like in the test-particle trajectories, the processes that establish the asymptotic energies and momenta appear to be stresses in the vicinity of the binary; once material expands significantly its properties are not dramatically altered.  

The sharp lower limit of the energy distribution corresponds to the energy of a given semi-major axis relative to the system's center of mass. The comparison to the ballistic trajectories reveals that the fluid settles into a configuration that overlaps in energy with some of the ballistic trajectories emitted from corotation near $L_2$. The typical gas angular momentum is significantly lower than the range of angular momenta occupied by the corotation free trajectories. This indicates that gas expands from the binary with less than corotation angular velocity (by approximately 20\%). 

The lower panel of Figure \ref{fig:bounddist}  plots the distribution of material in angular momentum-energy phase space. To produce this diagram, we select and plot only material with radius larger than that of the $L_2$ point relative to the system's center of mass ($r_{\rm com}>2.55R_1$). 
In the limit of vanishing enthalpy, there are contours of constant Jacobi parameter, $E_J$, in this plane because Jacobi parameter depends on the sum of kinetic plus potential energy along with angular momentum and orbital frequency, as described by equation \eqref{jacobi2}. 

The gas outflowing from the binary forms a relatively compact locus in energy-angular momentum phase space. The typical energy is, as suggested earlier, mildly negative, indicating that material is bound to the binary.  This distribution extends to mildly positive $\mathcal{B}$. A comparison to Figure \ref{fig:largescaleE} reveals that these regions lie in shock-heated layers close to the binary. 
Because the typical angular momentum is less than that of the $L_2$ corotation trajectories, material appears significantly offset in phases space.  Together these properties indicate that material occupies a Jacobi parameter that is larger (less negative) than any of the $L_2$-trajectories $E_{J,L_2}\approx -1.15$, with typical values for the simulated fluid closer to $E_J\approx -1.0$.

\subsubsection{Why is the Early Outflow Bound?}

Given that distributions of outflow energy and momentum begin broad but narrow with expansion, we have argued that interaction in the vicinity of the binary is crucial in establishing the outflow's asymptotic properties. Therefore, in order to understand the bound nature of the outflow,  we now study the gas dynamics of flow in the loss region.

\begin{figure*}[tbp]
\begin{center}
\includegraphics[width=0.99\textwidth]{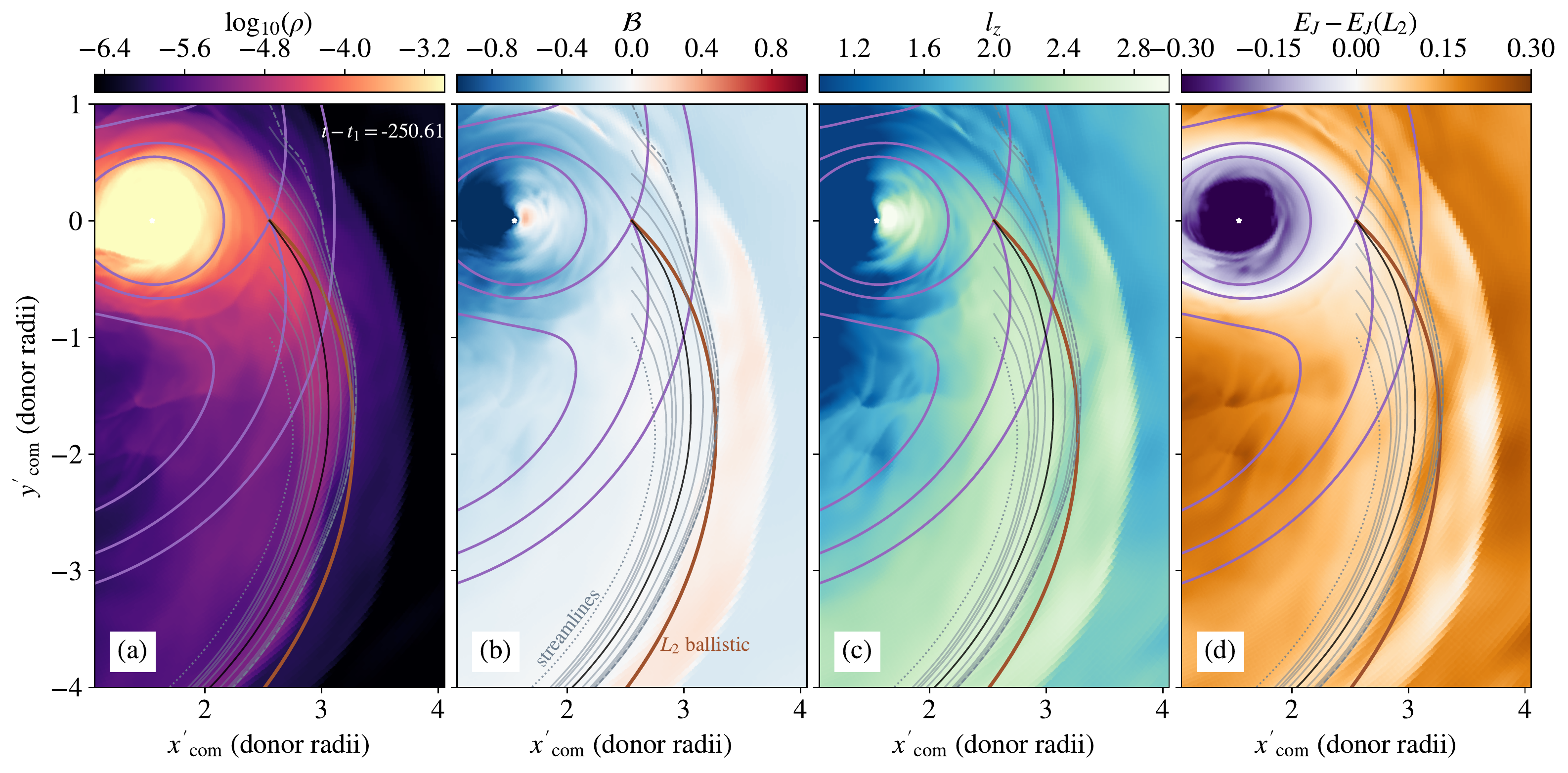}
\includegraphics[width=0.99\textwidth]{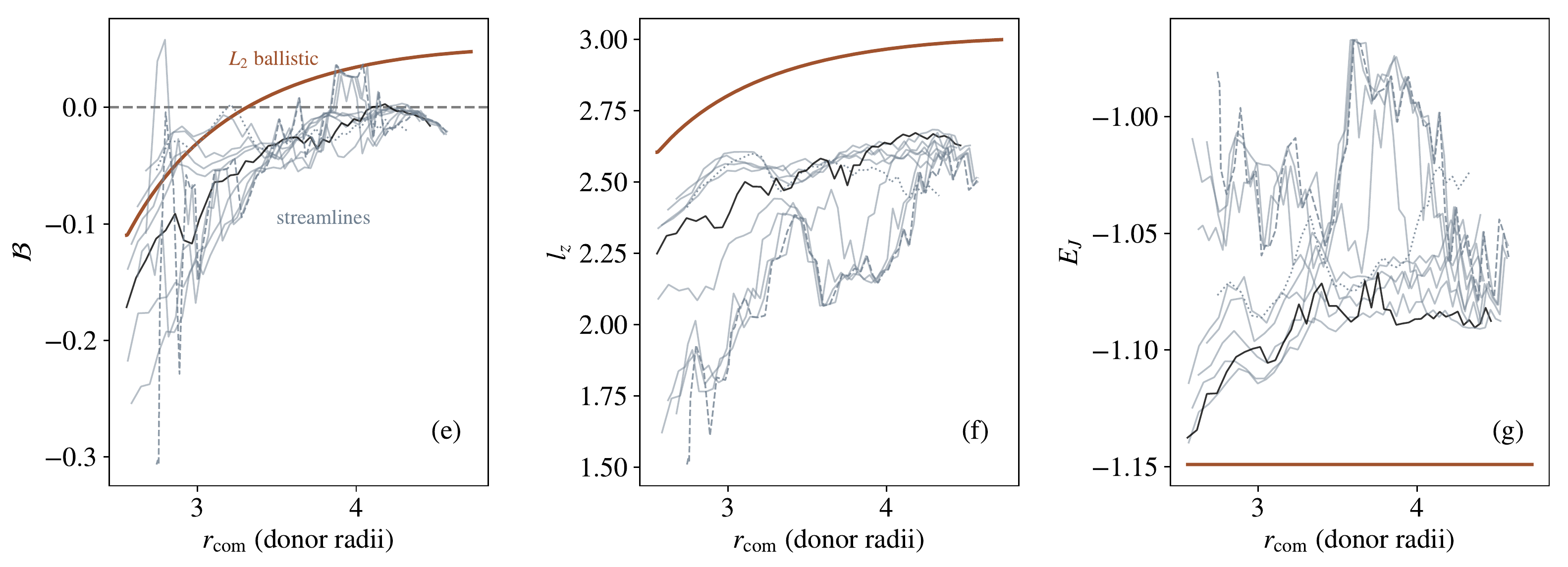}
\caption{The generation of a bound outflow at early time. Upper panels show fluid quantities on slices through the binary midplane at $t-t_1\approx-250$. Equipotential contours show the potentials that correspond to Lagrange points  $L_1$, $L_2$, and $L_3$. A thick brown line traces the path of a ballistic trajectory initialized from corotation at $L_2$. Finally, grey lines show instantaneous fluid streamlines initiated at a range of $y'_{\rm com}$ offsets relative to $L_2$. The lower panels trace fluid quantities along these streamlines. Relative to the ballistic trajectory, fluid streamlines expand more slowly from the binary, winding into a tighter spiral. They start with disparate positions, binding energies, and angular momenta, but converge into a narrow, relatively uniform stream.  A comparison with Figure \ref{fig:testparticle} reveals that this fluid interaction is forced by the convergence of trajectories in the binary potential -- this is the same region in which ballistic trajectories initialized above and below $L_2$ cross.
  }
\label{fig:boundstream}
\end{center}
\end{figure*}

Figure \ref{fig:boundstream} analyzes flow interactions with the outflow stream immediately following loss from the binary.  The upper panels show gas density, specific energy, specific angular momentum, and Jacobi parameter, respectively. In addition to equipotential contours and a ballistic trajectory initialized from corotation at $L_2$, we also plot a series of streamlines of the flow with thin grey lines, initialized at $\pm y'_{\rm com}$ positions relative to $L_2$. The lower panels of Figure \ref{fig:boundstream} sample the gas properties of energy, angular momentum, and energy as a function of distance from the center of mass along these streamlines, stopping where the exit the images in the upper panels.

The upper panels of Figure \ref{fig:boundstream} show that gas streamlines emanating from a broad region around the accretor and $L_2$ converge spatially. By the time the streamlines reach $y'_{\rm com}\approx-4$, a relatively concentrated stream has formed. Free trajectories initialized with similar offsets, as shown in Figure \ref{fig:testparticle}, cross after trailing away from the accretor at $y'_{\rm com}\approx-1$. 
Though the fluid trajectories cannot cross as is seen in the ballistic paths of Figure \ref{fig:testparticle}, compression from the tidal potential explains the narrowing of the  stream in this region.  
Finally, the fluid stream is more tightly-wound relative to the binary than the ballistic trajectory, indicating that it has lower radial velocity. 

Turning to the fluid properties sampled at the starting positions of the streamlines in Figure \ref{fig:boundstream}e,f,g,  we see that streamlines leave the vicinity of the binary with a wide range of specific energies and angular momenta. Of these streamlines, the one that is initially least bound starts on the trailing edge of $L_2$ ($-y'_{\rm com}$ offset). The remainder of the streamlines are significantly more-tightly bound to the binary at the outset, by up to a factor of three in total energy (See Figure \ref{fig:boundstream}e). 
For all of the sampled streamlines, the initial angular momentum is lower than that corresponding to corotation at $L_2$ (Figure \ref{fig:boundstream}f). 
The initial range spans roughly a factor of plus or minus 25\% around $l_z\approx2$. Initial Jacobi parameters, equation \eqref{jacobi1}, which are a function of both energy and angular momentum, also span a range, mostly occupying somewhat less-negative values than that of the $L_2$-corotation ballistic trajectory (Figure \ref{fig:boundstream}g). 
It is worth noting that in sampling gas that eventually outflows from the binary, we are capturing just the tip of a distribution that extends the bulk of the material which occupies much tighter binding energies and lower angular momenta within the system's Roche lobes.

Just as seen spatially, fluid properties converge within the stream as the gas expands away from the binary. Along with energy and momentum, gas Jacobi parameters also evolve with radius, converging toward an asymptotic value of approximately $-1.08$, as compared to a range of initial values spanning -1.14 to -1. Evolution of  gas Jacobi parameter along streamlines nicely quantifies the consequences of gas-dynamical interaction because, in the absence of pressure gradient forces, the Jacobi parameter is conserved along free trajectories. The convergence observed in stream properties can, therefore, be attributed to gas dynamical interaction that redistributes energy and angular momentum within a region where free-trajectories would intersect and cross. This gas-dynamical redistribution is responsible for shaping the narrow locus of the outflow in energy-angular momentum phase space as seen in Figure \ref{fig:bounddist}. 

As the gas expands relative to the binary, in addition to convergence of streamlines, the gas gains specific energy and specific angular momentum. A similar trend is seen in the ballistic trajectory and is due to continued acceleration by the binary potential. However, the asymptotic values of energy and angular momentum attained by material in the converging outflow stream are lower than those of the $L_2$-corotation trajectory (See Figure \ref{fig:boundstream}e,f). Because much of the outflow is initially more tightly bound and has lower angular momentum than corotation at $L_2$ would imply, when gaseous interaction averages the properties of outflowing material, the net energies and momenta achieved are lower than would be expected given corotation at $L_2$ as a starting condition. As gas continues to expand outside the loss region discussed here, we find that similar self-interactions continue to be important in decelerating equatorial fluid relative to comparable ballistic trajectories.  

Taken together, these features reveal that the dynamics of Roche lobe overflow from the donor toward the accretor end up populating a wide range of specific energies and angular momenta relative to the binary.  In particular, as flow around the accretor self-intersects a counter-rotating surface layer forms and eventually supplies material lost near $L_2$. As a result, material crosses the outer Lagrange point with less than corotation energy or angular momentum.  As fluid flow converges outside of $L_2$, gas properties average and the asymptotic properties of the outflow are lower-energy and angular momentum than would be predicted by a ballistic outflow from $L_2$. The qualitative change here is very significant, however, because this implies that instead of acquiring mildly-positive energy, gas total energy remains negative and the fluid piles up into a circumbinary torus.   Lacking a mechanism to hold material in corotation near $L_2$ (as might, for example, be the case in a contact binary in which both stars completely filled their Roche lobes) we expect these findings to be robust with respect to variations in specifics, like mass transfer rate or gas sound speed.

\subsection{Unbound Ejecta Phase}\label{sec:unbound}

Unbound ejecta are generated in substantial quantities only late in the  evolution toward binary coalescence, as the accretor skims the surface of the donor, at separations less than roughly $1.5R_1$ (see Figures \ref{fig:masslosstime} and \ref{fig:masslosssep}). In Figure \ref{fig:largescaleE} these late-stage ejecta reveal themselves as spiral trace of unbound material from roughly the half-orbit prior to the plunge of the accretor within the donor's envelope. 
 
\subsubsection{Flow Morphology}

\begin{figure*}[tbp]
\begin{center}
\includegraphics[width=0.38\textwidth]{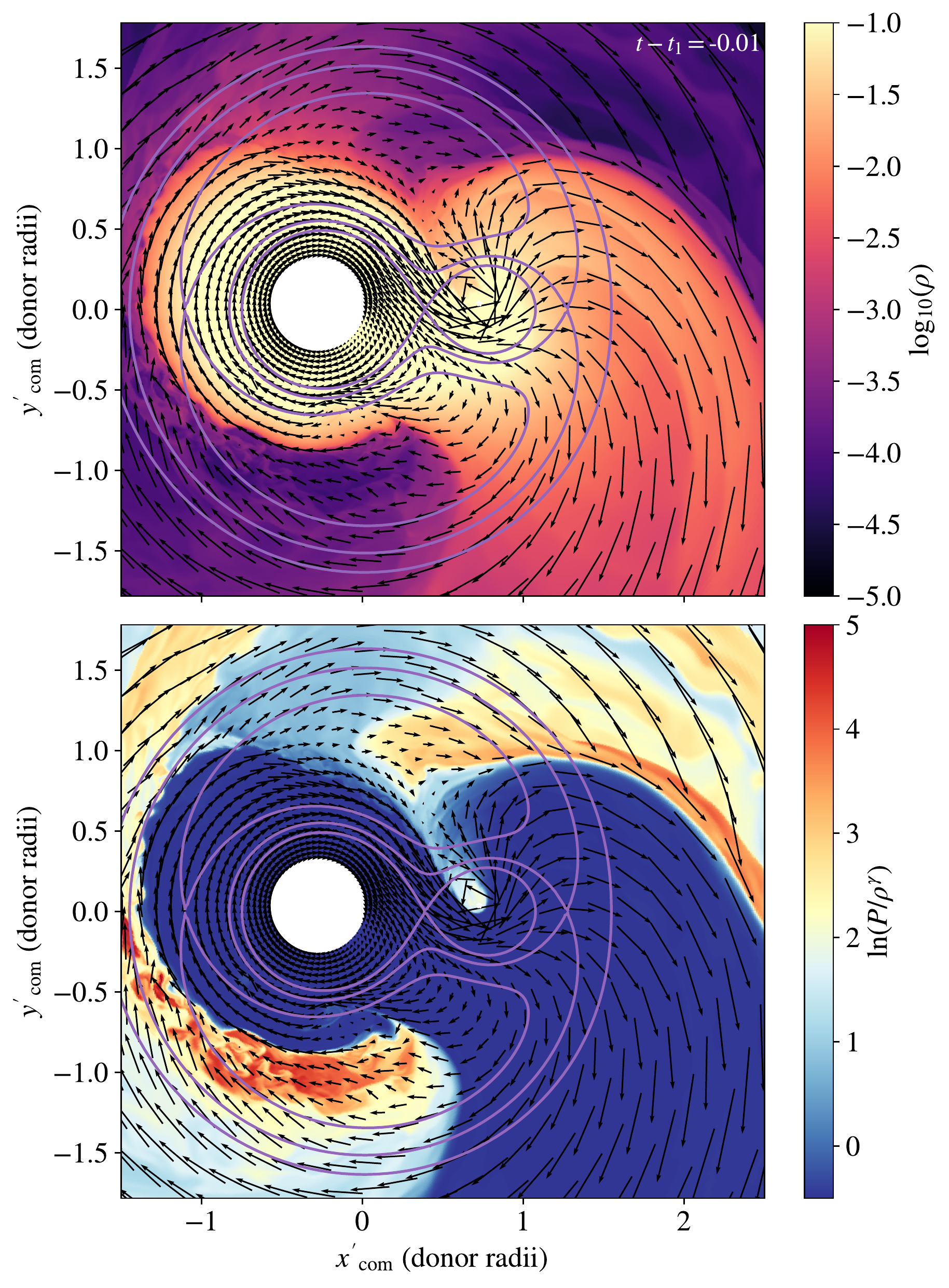}
\hspace{1cm}
\includegraphics[width=0.55\textwidth]{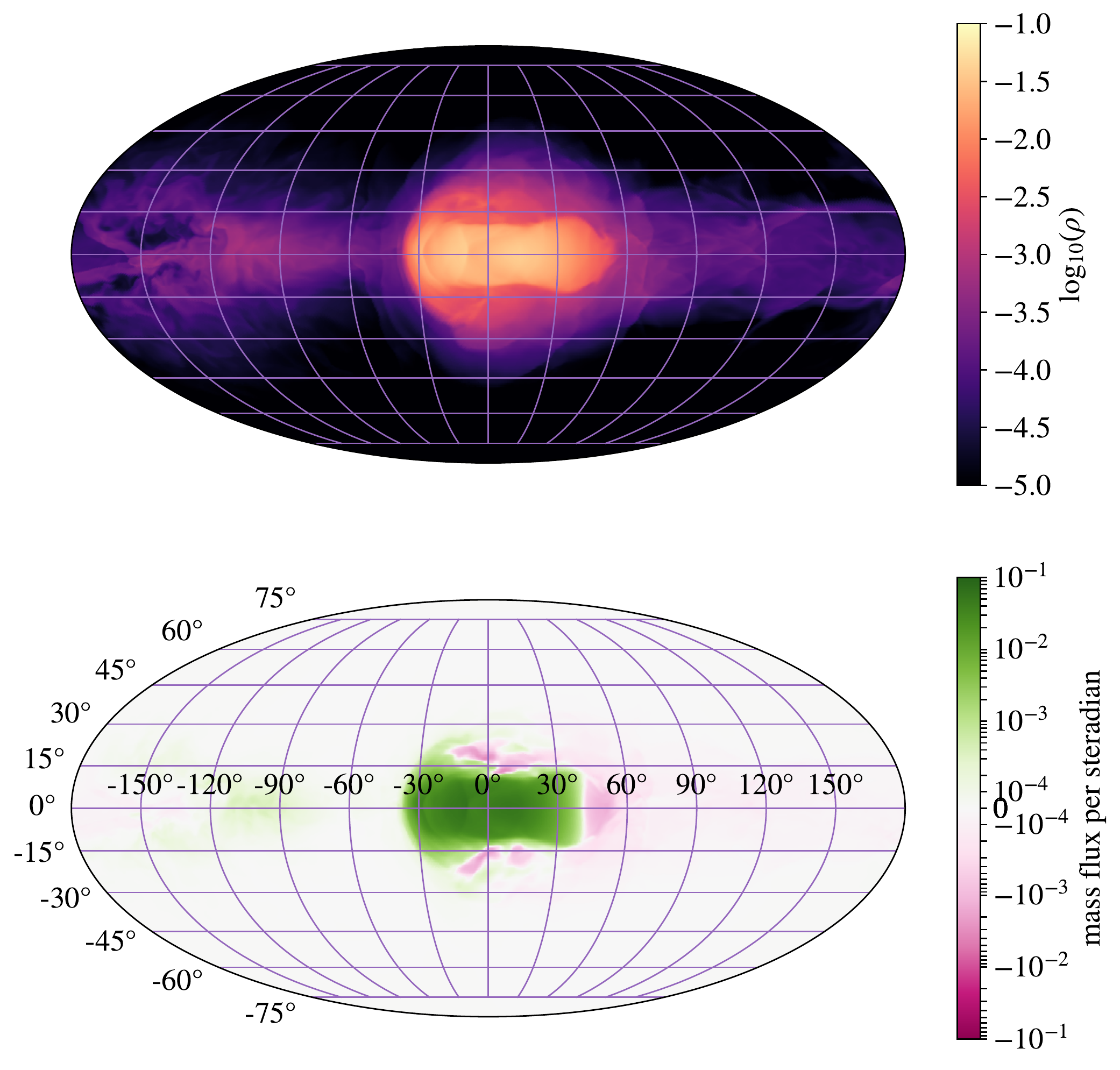}
\caption{ Same as Figure \ref{fig:binaryflow0}, but now showing a snapshot from near the moment of binary coalescence, when the separation is approximately the donor star's original radius.  The outflow morphology has changed significantly, with a broad fan of low-entropy ejecta arcing outward from the vicinity of the accretor. In the spherical slices (relative to the donor, at the radius of the $L_2$ point), this fan of ejecta appears as by far the strongest area of outflow. We see relatively uniform outward mass flux spanning roughly $\pm 35$~degrees in azimuth and $\pm 15$~degrees in altitude relative to the accretor.  }
\label{fig:binaryflow1}
\end{center}
\end{figure*}

A distinct flow morphology in the vicinity of the binary emerges as the two components spiral closer. Figure \ref{fig:binaryflow1} portrays the gas flow properties, analogous to Figure \ref{fig:binaryflow0} except  near the moment of merger when the binary separation is equal to the original donor-star radius. 

The left panels of Figure \ref{fig:binaryflow1}, which slice through the binary midplane, reveal a wide stream of material being pulled from deep within the donor by the accretor's gravitational influence. This gas emanates from the binary in a broad fan \citep[e.g.][]{2008ApJ...672L..41R, 2014ApJ...786...39N,2016ApJ...816L...9O,2017MNRAS.464.4028I,2018ApJ...863....5M}. Whereas, earlier in the coalescence, we saw a disk around the accretor, at late times, we observe no such feature. Instead, material is relatively unconfined by the Roche lobes. One reason is the extreme degree of overflow of the donor outside its Roche lobe. Another is that the donor material retains nearly its original rotation. By the time the binary orbit tightens, and the orbital frequency increases correspondingly, the donor's envelope no longer rotates synchronously with the orbital motion \citep[][Figure 16]{2018ApJ...863....5M}. 

A comparison of gas entropy in Figure \ref{fig:binaryflow1} to the earlier snapshot of Figure \ref{fig:binaryflow0} is also revealing. In this later snapshot, material flung into the tails has entropy similar to that of the donor's envelope, indicating that it is not strongly shocked at this stage. Boundary layers between colliding flow regions form the only areas of higher-entropy material in Figure \ref{fig:binaryflow1}.  Earlier in the inspiral, only higher-entropy material escapes the binary when driven out of the Roche lobes by gas pressure gradients. 

The right-hand maps of Figure \ref{fig:binaryflow1} show density and mass flux on a spherical surface intersecting the $L_2$ point about the donor ($r=1.50R_1$). We note that the density and mass flux scales are shifted several orders of magnitude higher (while still showing the same dynamic range) as compared to the corresponding panels of Figure \ref{fig:binaryflow0}. The highest density gas in this surface is found in the fan of ejecta, which remains concentrated near the equator. The primary distinction as compared to the earlier snapshot of Figure \ref{fig:binaryflow0} is the broad, nearly uniform mass-flux outflow region that extends approximately $\pm$35 degrees in azimuthal angle relative to the accretor (whereas Figure \ref{fig:binaryflow0} exhibits a concentration in the approximate vicinity of $L_2$). 

A final feature of note at this stage is that the properties of the binary and the ejecta are rapidly evolving. Whereas the flow early in the coalescence is in a roughly steady-state, now the binary orbit is plunging together on a timescale similar to the donor's dynamical time, as shown in Figure \ref{fig:masslosstime}. The implications of this for outflow from the binary are that the binary is both tightening and increasing in orbital frequency as the ejecta expand. As seen in Figure \ref{fig:largescaleE}, unbound ejecta have not yet propagated very far from the converging binary in this, later snapshot. 

\subsubsection{Unbound Ejecta}

\begin{figure}[tbp]
\begin{center}
\includegraphics[width=0.49\textwidth]{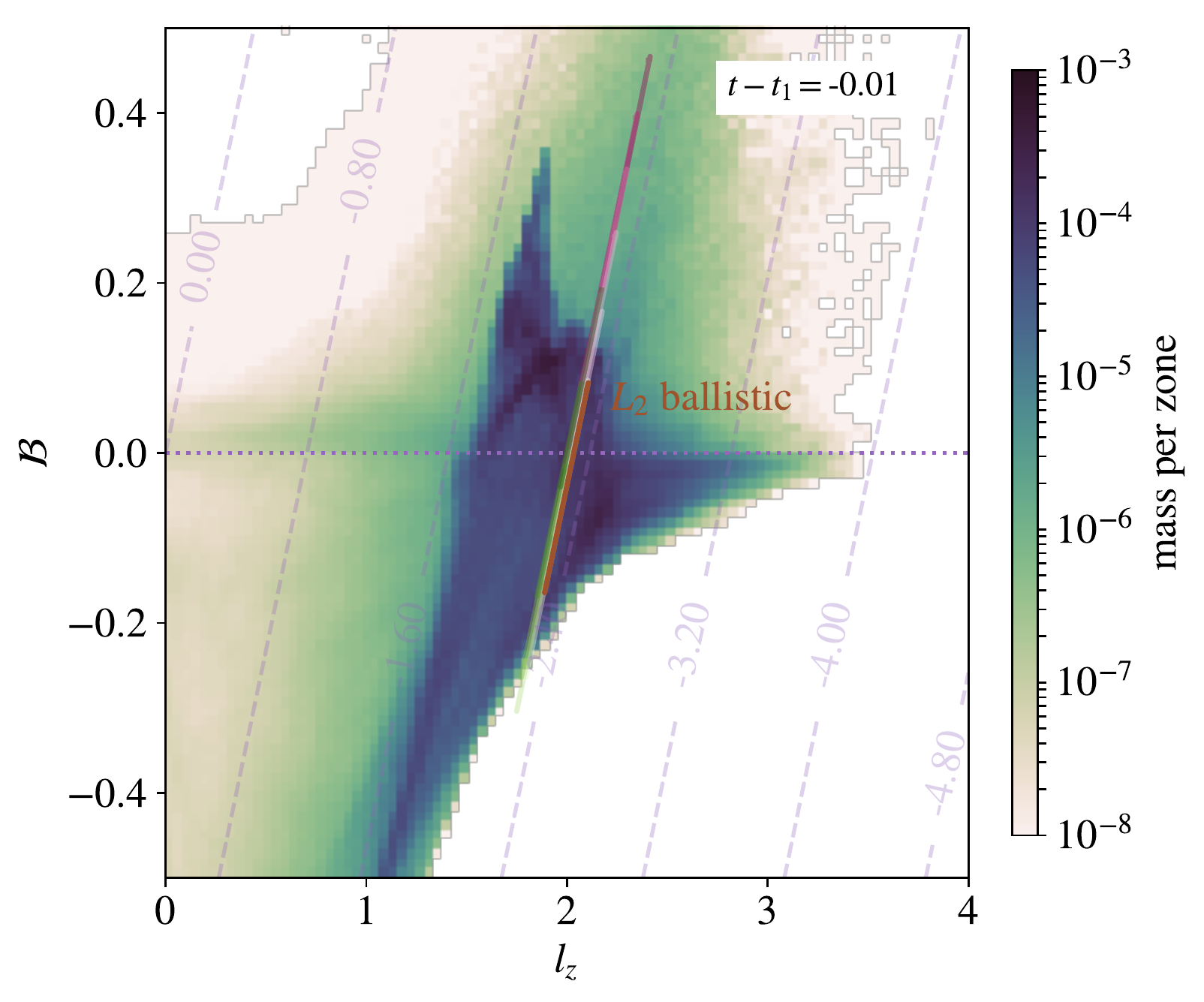}
\caption{ The energy-angular momentum phase space of ejecta beyond the radius of $L_2$ at late time. In contrast to Figure \ref{fig:bounddist}, we now observe a very broad distribution in energy that extends significantly into positive energies (or unbound trajectories).   }
\label{fig:unbounddist}
\end{center}
\end{figure}

Unbound ejecta, generated during the late binary coalescence, populate a different energy-angular momentum phase space than that of the circumbinary torus material that is lost earlier. We analyze the distribution of these late-stage ejecta in Figure \ref{fig:unbounddist}. Just as in Figure \ref{fig:bounddist}, here we select material outside of the radius of $L_2$. Contours in the background correspond to fixed Jacobi parameter in the limit of vanishing internal energy, and the ballistic trajectories of test particles, initialized from corotation near $L_2$, are again shown with colored lines. 

The full morphology of Figure \ref{fig:unbounddist} is complex, particularly because this represents the instantaneous overlay of material lost during various moments in the binary's rapidly-evolving coalescence. The slices of Figure \ref{fig:binaryflow1} reveal that the initial properties of ejecta are very different from that of our example ballistic trajectories. Rather than starting nearly at rest in the system's corotating frame, gas is flung outward with a wide range of radial and angular velocities. Given these drastic flow differences, it is, perhaps, unsurprising that the gas distribution spreads much more broadly, particularly in angular momentum space, than the example ballistic trajectories. 

Nonetheless, a comparison to the earlier distribution of Figure \ref{fig:bounddist} is illuminating. There remains some overlap in the distributions, especially at negative total energies and $l_z\approx2.5$. This represents material lost prior to the last few dynamical times of plunging inspiral. However, compared to the previous outflow, the full distribution now spreads across a much wider range of specific energies and angular momenta.  In particular, a new, lower angular momentum component of outflow populates a track of Jacobi parameter between $-1.6$ and $-2.0$ and extends to energies that are substantially larger then zero. This new component represents the unbound ejecta that are forming as the binary coalesces. 

\subsubsection{Why are the Late Ejecta Unbound?}

\begin{figure*}[tbp]
\begin{center}
\includegraphics[width=0.99\textwidth]{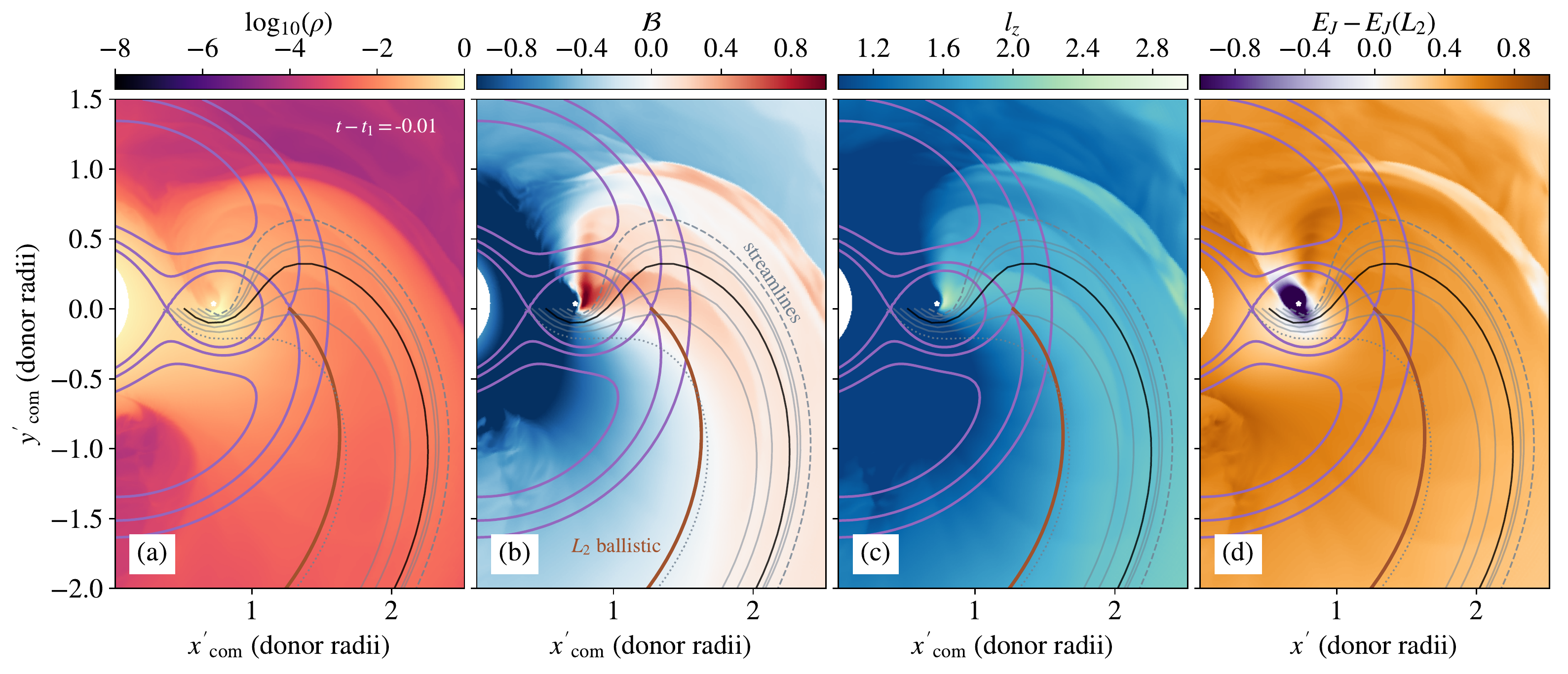}
\includegraphics[width=0.99\textwidth]{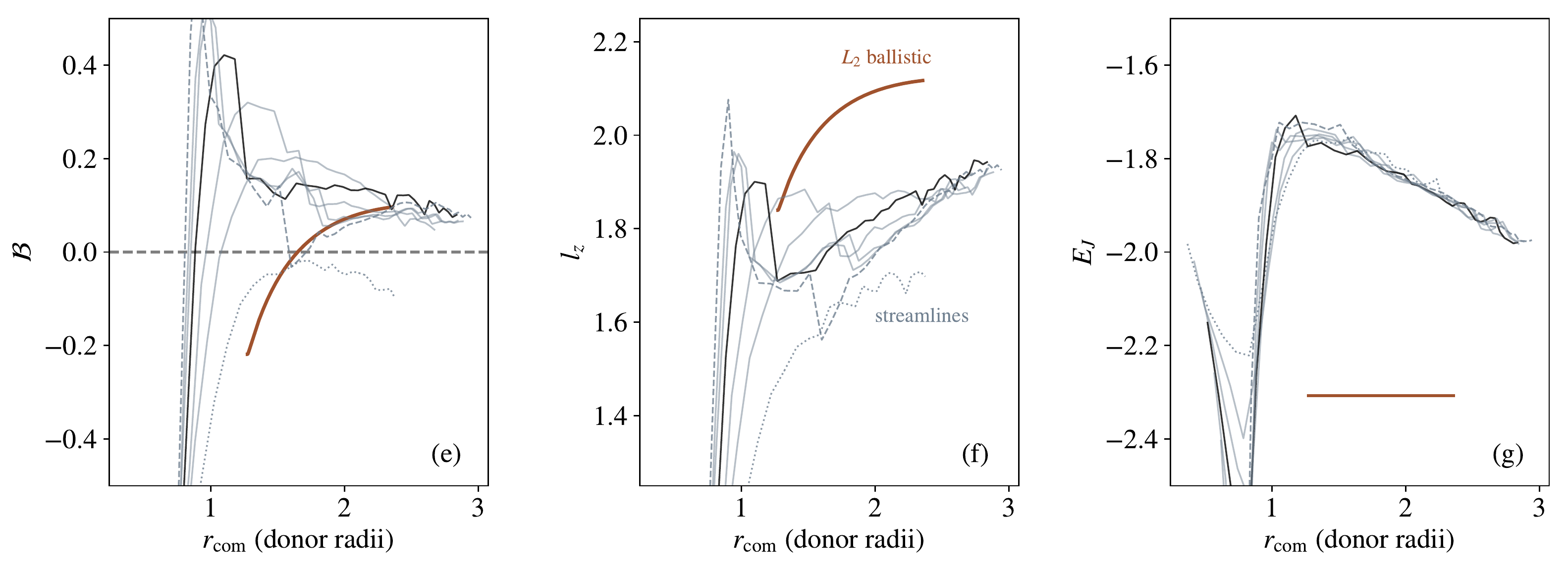}
\caption{ The formation of unbound ejecta at late time. Identical to Figure \ref{fig:boundstream}, but now showing the snapshot nearer to coalescence. Streamlines begin deep in the donor's envelope and arc around the accretor, deflected by its gravitational influence, and are flung outward. Compared to the ballistic trajectory initialized from corotation at $L_2$, the streamlines have significantly more radial velocity, but less angular momentum. This leads much of the ejecta to achieve positive total energy just after passing the accretor. Jacobi parameters of the ejecta are significantly offset from that of the $L_2$-corotation free trajectory, their quantitative properties are discussed further in the text.
}
\label{fig:unboundstream}
\end{center}
\end{figure*}

Having identified the presence of unbound ejecta late in the binary coalescence, here we analyze its connection to the dramatic change in outflow morphology that is observed at these stages. We analyze this later phase of coalescence in Figure \ref{fig:unboundstream}. 

Much of Figure \ref{fig:unboundstream} is analogous to Figure \ref{fig:boundstream}, but here we initialize streamlines inside the donor's envelope in order to trace the gravitational slingshot they receive from the accretor. The specific energy panel shows that much of the broad fan of ejecta that is generated is energetically unbound immediately upon passage by, and gravitational interaction with, the secondary. Compared to the ballistic trajectory initialized in corotation at $L_2$, most of the fluid streamlines initially expand more rapidly away from the binary system, indicating that they have a larger radial component of their velocity. 

Tracing quantities along streamlines in Figure \ref{fig:unboundstream}, we see from panels  panels (e) to (g) that fluid begins with a wide range of energies and angular momenta that converge as the gas expands, much like seen in Figure \ref{fig:boundstream} at earlier times. The asymptotic energy is positive, and happens to be similar to that of the $L_2$ ballistic flow despite the very different ejecta geometry. The asymptotic angular momentum is significantly lower than that of the $L_2$ ballistic trajectory. This occurs because most gas never reaches corotation with the $L_2$ point and instead has a larger radial, but lower azimuthal, velocity upon ejection. This combination of energies and angular momenta generate typical fluid Jacobi parameters that are significantly offset relative to the corotation trajectory from $L_2$ (note the broader color-scale in Figure \ref{fig:unboundstream}d, as compared to Figure \ref{fig:boundstream}d). 

Unlike the streamlines sampled in the earlier snapshot of  Figure \ref{fig:boundstream}, the Jacobi parameters of the fluid  are tightly clustered at a given radius, but evolve significantly with expansion in $r_{\rm com}$ (Figure \ref{fig:unboundstream}d and \ref{fig:unboundstream}g). We see a major increase in Jacobi parameter as gas passes by the accretor, indicating that gaseous stresses (pressure gradients) are modifying the flow trajectories substantially in this region. This impression is confirmed by comparison of the streamlines shown to ballistic trajectories initialized with identical positions and velocities. 

Quantitatively, the magnitude of the maximum Jacobi parameter reached, roughly $E_J\approx-1.8$ at $r_{\rm com}\approx1$, can be understood from equation \eqref{jacobi2} and the energy and angular momentum of the streamlines shown in Figures \ref{fig:unboundstream}e.  Approximately, the energy is near zero and the orbital frequency is near unity at this separation; as a result, we find $E_J \sim - l_z$.   An important caveat is that the  slope of decreasing $E_J$  at $r_{\rm com} \gtrsim1$, does not represent gaseous stress and is, instead, an artifact of the rapidly evolving binary orbit. The Jacobi parameter's definition is useful in the context of static binary properties. However, given the rapid orbital tightening seen in Figure \ref{fig:masslosstime} at this stage, material at larger radii was ejected when the binary had different properties than material at smaller radii.

Two key factors drive the dramatic change observed in flow properties described above -- and the accompanying generation of unbound ejecta. One of these factors is that the degree of Roche lobe overflow becomes increasingly extreme as the binary trends toward coalescence. This is of significance because it implies that there is donor material at a range of (smaller) impact parameters relative to the accretor. This is seen perhaps most clearly in the left panels of Figure \ref{fig:binaryflow1} and in the upper-left panel of Figure \ref{fig:unboundstream}.  A second key factor is that orbital motion desynchronizes from envelope rotation as the two stars coalesce. In \citet{2018ApJ...863....5M}, we demonstrated that the donor maintains approximately constant rotation even as the orbital frequency increases dramatically from the Roche limit to engulfment. 

The combination of these two factors implies that when $a\approx R_1$, the accretor is skimming the surface of the donor star with relative velocity similar to the orbital velocity. In the instantaneously-corotating frame, gas in the donor's envelope appears counter-rotating, as the velocity vectors of Figure \ref{fig:binaryflow1} show.  In this situation, gas at low impact parameter relative to the accretor receives the largest angle scattering and is deflected significantly away from its original trajectory by gravitational interaction with the accretor.  A range of impact parameters, as sampled by the streamlines of Figure \ref{fig:unboundstream}, generate a range of scattering angles and resultant trajectories away from the binary. 

In the frame of the accretor, the ``headwind" of stellar material spans a very steep gradient in density -- that of the stellar limb. As a result, dense material focussed from the stellar envelope, compressed, and deflected outward is nearly uninhibited by the comparatively low density stellar atmosphere it is redirected into. \citet{2015ApJ...803...41M,2015ApJ...798L..19M,2017ApJ...838...56M,2017ApJ...845..173M} performed hydrodynamic simulations of flow in the immediate region around an accretor embedded within the envelope of its companion, and predicted similar, very asymmetric flow morphologies in cases of extreme density gradient across the gravitationally-focused material. As the streamlines of Figure \ref{fig:unboundstream} show, the portion of these ejecta most-strongly redirected by this flow asymmetry  become unbound almost immediately after passing the accretor.

\section{Bipolar Collimation}

We have seen that higher-velocity unbound ejecta  follow lower-velocity bound outflow. These components are destined to interact. Here we describe how this interaction redirects material toward the system's poles. 

\subsection{Ejecta-Outflow Interaction}

\begin{figure}[tbp]
\begin{center}
\includegraphics[width=0.49\textwidth]{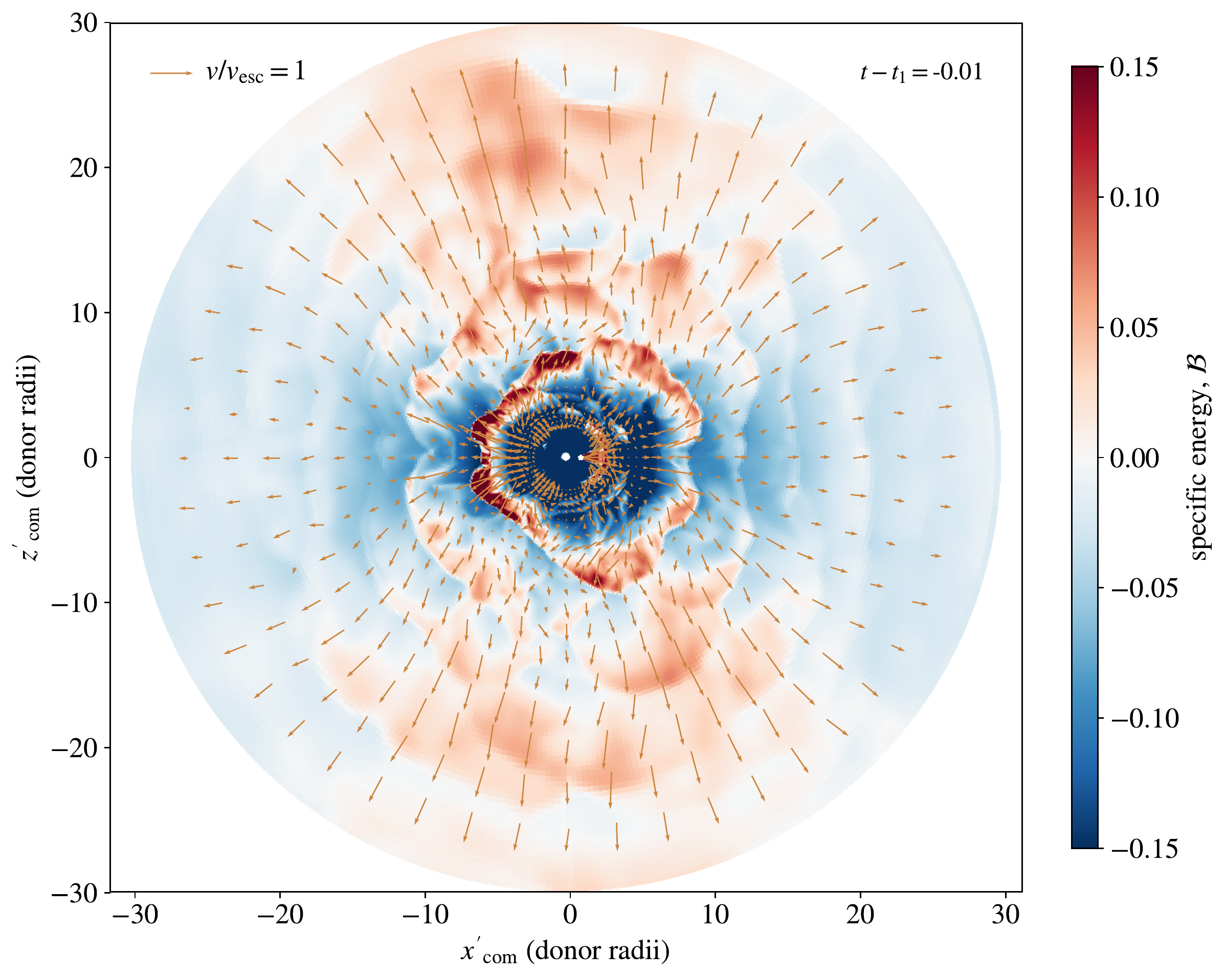}
\includegraphics[width=0.49\textwidth]{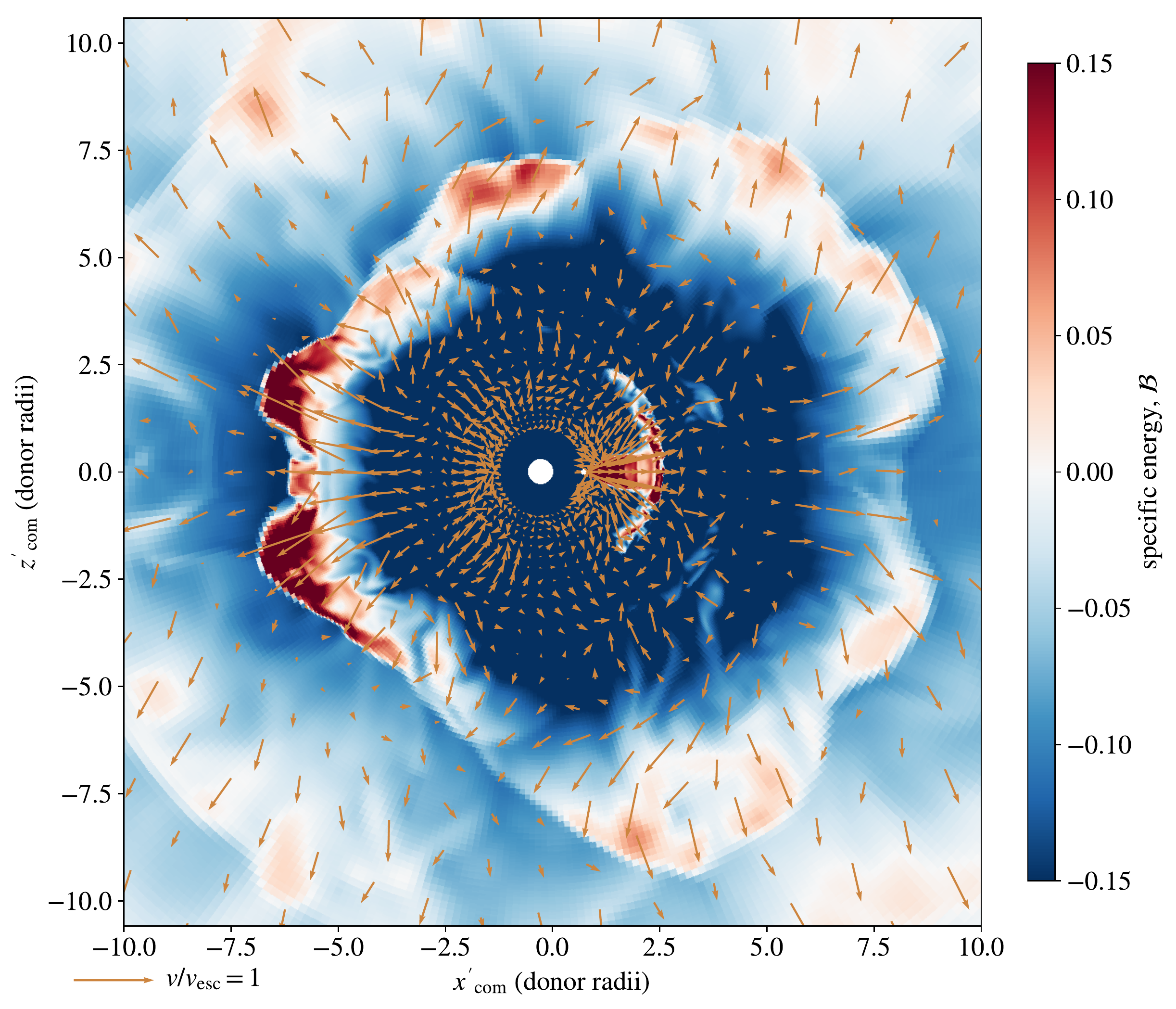}
\caption{The formation of bipolar outflows at time near coalescence through the redirection of equatorial ejecta. The color scale shows slices of gas specific binding energy perpendicular to the orbital midplane. Vectors show gas velocities rescaled to the local escape velocity. Unbound ejecta originate in the equatorial plane where they interact with previously-lost bound gas, which occupies an equatorial torus. Ejecta are decelerated by this interaction, but also diverted increasingly toward the poles such that outside of roughly ten donor radii, unbound material only exists in two broad polar cones. 
}
\label{fig:Evertical}
\end{center}
\end{figure}

\begin{figure*}[tbp]
\begin{center}
\includegraphics[width=0.99\textwidth]{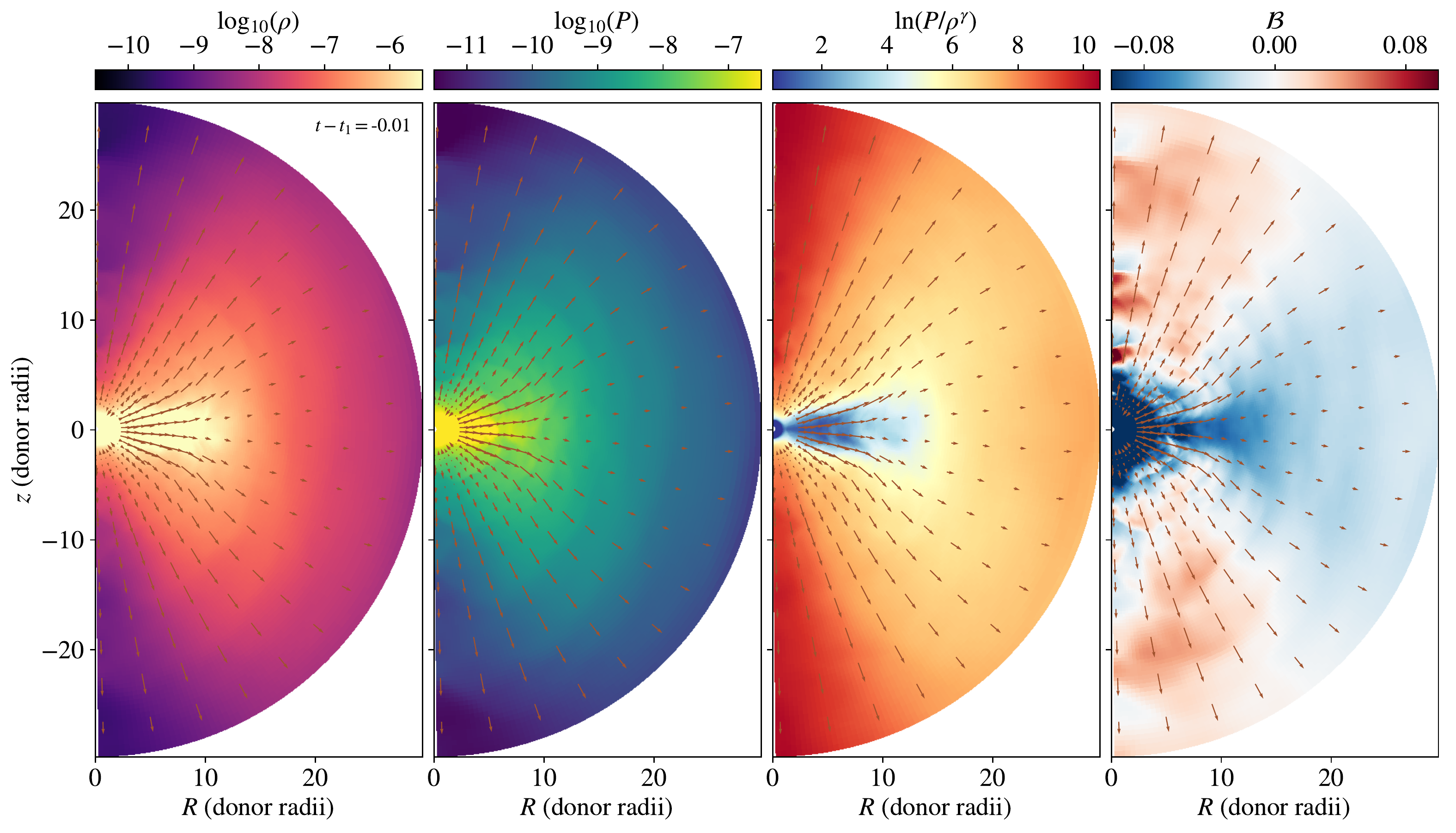}
\caption{Late-time azimuthal averages (relative to the donor star) of gas density, entropy, and total energy, with azimuthally-averaged velocity field also plotted. The equatorial overdensity of bound gas (previously ejected) results in shocks that redirect ejecta toward the poles. Unbound material and the largest outflow velocities are found within roughly 40 degrees of the poles.  }
\label{fig:azimuthalavg}
\end{center}
\end{figure*}

Early outflows of bound material in the equatorial plane of the binary leave a torus of bound gas surrounding the ongoing merger (see Figure \ref{fig:largescale}, for example). As new outflows and ejecta exit the system with increasing radial velocity, the leading edge of newly-lost material interacts strongly with the trailing edge of previously-lost gas. 
These interactions happen continuously in the spiral shocks that dominate the morphology of the debris in the orbital plane (see Figure \ref{fig:largescaleE}). These gaseous interactions redirect material away from the orbital plane as it seeks a path of less obstruction.  Increasingly, these interactions slow the radial velocity of gas in the orbital, $x-y$, plane while directing an increasing component toward the poles, in the $\pm z$-directions. 

Figure \ref{fig:Evertical} shows the distribution of gas specific binding energy in the vertical plane at the time of engulfment of the accretor. In this plane, the spiral features of consecutive orbits of ejecta (Figure \ref{fig:largescaleE}) make alternating appearances at positive and negative $x'_{\rm com}$. Near the accretor, unbound ejecta are originally present along the plane of the orbit -- analogous to what was observed in the maps of Figure \ref{fig:binaryflow1}. Interaction with the surrounding, bound material broadens the distributions of unbound material and reshapes the outflow toward the polar directions (as seen in the zoomed-in panel of Figure \ref{fig:Evertical}).  As we step outward on alternating sides of the binary tracing the spiral wave of ejecta, we see that material in the midplane is slowed most severely, eventually to the point of having negative specific energy. Unbound gas is increasingly found exclusively above and below the midplane. 

Zooming out to the broader view of the upper panel of Figure \ref{fig:Evertical}, we see that shocks from alternating sides of the binary converge near the poles and accelerate material in the $\pm z$-directions. The velocity arrows, scaled to the approximate local escape velocity, $\sqrt{G(M_1+M_2)/r_{\rm com}}$, show that large equatorial velocities are redirected to large polar radial velocities within about 20 donor-star radii. At the edges of the simulation domain, we observe that unbound gas escapes only along wide, conical outflows in the polar directions.

\subsection{Conical Outflows}

In Figure \ref{fig:azimuthalavg}, we examine the angle-averaged  morphology that emerges from this interaction. We azimuthally average gas density, pressure, specific entropy, and specific energy about the donor star. We see a dense, bound torus with a strong equatorial concentration, especially at small radii. The highest pressures and lowest specific entropy material lies within the inner parts of this torus, in the binary midplane. At larger radii in the midplane, higher specific entropies result from the spiral shocks. The resultant vertical pressure gradients redirect ejecta gas away from the bound torus and toward the poles. 

Velocity vectors reflect azimuthal averages as well. We see that at small radii there are generally large radial velocities in the equatorial midplane, but at slightly larger separations, $R\gtrsim10$, radial velocities are low (but positive) in the bound torus of material that opens about the binary midplane.  
 We might wonder why energetically-bound gas does not entirely stall or fall back toward the binary, leading to negative radial velocities. While this would occur given a clear environment, the continued impingement of newly-ejected material prevents significant fallback at these stages \citep[see a similar discussion in section 4.4 of ][]{2018arXiv180902297R}. 

Moving toward the poles, we see lower densities and higher specific entropies. These properties are accompanied by higher radial velocities at large radii and positive $\mathcal{B}$. The bipolar cones of relatively fast-moving but low-density ejecta that emerge in this particular case have opening angles of approximately 40 degrees, though we strongly suspect that the quantitative value is subject to variation under different conditions.

\begin{figure}[tbp]
\begin{center}
\includegraphics[width=0.485\textwidth]{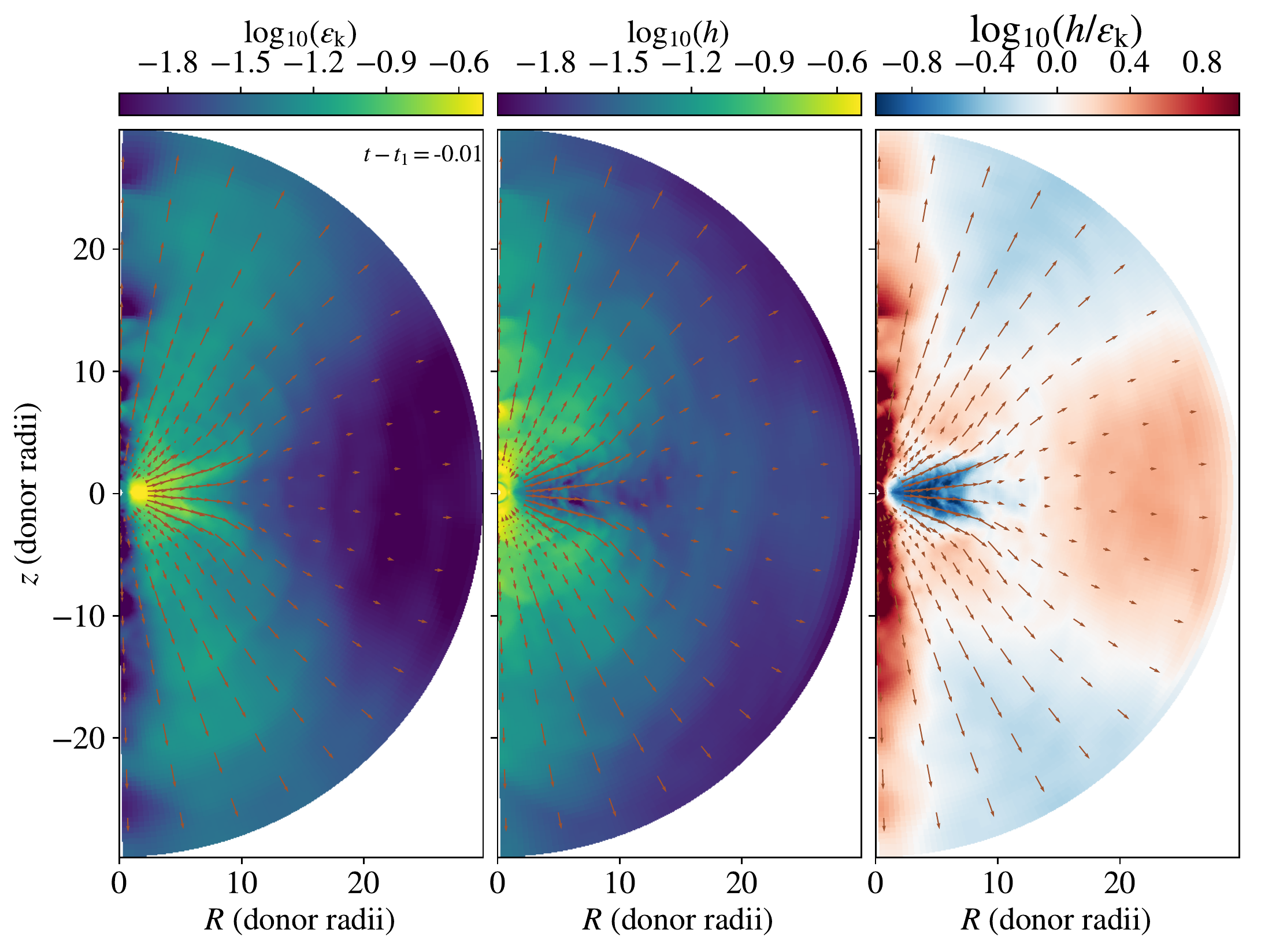}
\caption{  Late-time azimuthal averages of gas specific kinetic energy, specific enthalpy, and their ratio. Material is ejected with high kinetic energy content near the equator, but becomes thermally-dominated as it stalls. Conical outflows, at approximately 40 degrees from either pole, originate with the thermal energy from ejecta self-interaction and accelerate to ballistic outflow prior to exiting the simulation domain. The strong feature in $h/\varepsilon_{\rm k}$ near the poles is a remnant of the thermally-dominated initial conditions of the calculation.   }
\label{fig:azimuthalavg_kinematics}
\end{center}
\end{figure}

Figure \ref{fig:azimuthalavg_kinematics} examines the kinematics and energetics of the conical outflows. In particular, we compare the gas specific kinetic energy to the specific enthalpy. These terms, represent, roughly the breakdown of gas energy into kinetic and thermal components, see equation \eqref{bernoulli}. Finally, the right-hand panel of Figure  \ref{fig:azimuthalavg_kinematics} displays the ratio $h/\varepsilon_{\rm k}$. The regions of highest specific kinetic energy are close to the donor in the plane of the orbit. These features represent the original ejection of material from the binary, as discussed in Section \ref{sec:unbound}. At larger radii, specific kinetic energy is highest within the conical outflow, not in the equator. Gas specific enthalpies, on the other hand, are highest close to the binary, but out of the orbital plane. High thermal energy content in these regions is produced by the shocks that direct material out of the plane of the orbit (Figure \ref{fig:Evertical}). 

The ratio $h/\varepsilon_{\rm k}$ reveals that ejecta in the midplane begin with mostly kinetic energy, which transitions to mostly thermal energy as they are decelerated by the surrounding torus material. Within the conical outflows, material is thermally-dominated at smaller radii ($R\lesssim 10$), but are accelerated by pressure gradients to kinetic energy-dominated (or largely-ballistic) outflows at larger radii. Figure \ref{fig:azimuthalavg} shows that within the bi-conical outflow, $\mathcal B$ is nearly constant along the radial direction, indicating that this transition occurs at constant Bernoulli parameter. Very near the system pole, $h\gg \varepsilon_{\rm k}$. These are regions of very low density that remain dominated by the initial, high entropy background (compare to Figure \ref{fig:azimuthalavg}).

Bipolar ejecta directly results from the mechanical collimation of initially-equatorial ejecta into the polar directions. The collimation is enabled by the emission of bound outflows in the early stages of binary coalescence. Were it not for the bound torus of early outflow, we would instead expect a largely planar sheet of unbound equatorial ejecta at late times. 

\subsection{Comparison to Related Studies}

\citet{2006MNRAS.365....2M} predicted bi-conical outflows from binary mergers, motivated by models in which they manually added first angular momentum, then energy to a gaseous sphere.  This is qualitatively consistent with launching an explosion inside a rotationally-distorted envelope. While these numerical models were clearly crude, their predicted angular outflow distributions \citep[see, for example,][Figure 3]{2006MNRAS.365....2M} have substantial qualitative similarity to those seen in Figure \ref{fig:azimuthalavg}. Subsequently, \citet{2007Sci...315.1103M,2009MNRAS.399..515M} applied their findings to the formation of the nebular remnant of supernova 1987A.

More recently, several numerical experiments have examined similar gas dynamics in the context of common-envelope binary systems \citep{2017ApJ...850...59P,2017MNRAS.471.3200M,2018ApJ...860...19G,2018arXiv180705925F,2018arXiv180902297R}. Each of these works examined the interaction of a dense torus with fast, spherical ejecta. \citet{2017ApJ...850...59P,2017MNRAS.471.3200M} considered the case of pre-merger mass loss from $L_2$ forming a circumbinary torus, as we have described here, followed by spherical ejecta from the moment of merger. \citet{2018ApJ...860...19G,2018arXiv180705925F,2018arXiv180902297R} simulate the interaction of a post-merger circumbinary torus with wind from the perturbed remnant star. Each of these experiments find that flow escapes along the polar axes, generating extremely aspherical geometries \citep[see, for example, Figure 13 of][]{2018arXiv180902297R}. Compared to these scenarios, our finding is somewhat more extreme -- we observe that ejecta from the moment of merger are equatorial, not spherical; even so, these ejecta are redirected into conical outflows near the poles.

\section{Conclusions}

Evidence for bipolar remnants of stellar coalescence from recent observations \citep{2018arXiv180401610K}, outlined in Section \ref{sec:bipolarobsv}, prompts many questions about the physical origin of these features within a stellar merger or common-envelope interaction. The shaping and kinematics of these remnants are key properties that we would like to understand in order to decode the physical processes that eject mass and transform binaries during common envelope phases. 

In this paper, we simulate the early stages of binary coalescence leading to a common envelope phase -- from the onset of unstable mass transfer with Roche lobe overflow until the accretor star becomes engulfed within the envelope of the more-extended donor star.  Some key findings of our study are:

\begin{enumerate}

\item The lead-in to stellar coalescence is characterized by extensive outflows from the binary system that arise when material transferred from the donor toward the accretor then overflows the accretor's Roche lobe near $L_2$ (Figure \ref{fig:largescale}).  Orbital decay and cumulative mass loss from the donor to the circumbinary environment are correlated and accelerate as the binary separation tightens (Figures \ref{fig:masslosstime} and \ref{fig:masslosssep}).

\item In contrast to previous predictions, early outflows from the coalescing binary remain bound and build up in the circumbinary environment into a thick torus with characteristic size of tens of donor-star radii. Later in the coalescence, unbound ejecta form (Figures \ref{fig:largescale} through \ref{fig:largescaleE}).

\item Previous predictions for unbound mass loss from binaries  \citep{1979ApJ...229..223S,2016MNRAS.455.4351P,2016MNRAS.461.2527P,2017ApJ...850...59P} are based on the assumption that gas flows leave the binary from corotation at the outer Lagrange point near the accretor, $L_2$. Our simulations show (Figure \ref{fig:binaryflow0}) that the early ejection of gas indeed occurs near L2.  However, gas carries a range of specific energies and angular momenta, most lower than that of corotation at $L_2$. As material expands, energy and angular momenta redistribute among the gas such that the eventual net properties are of a mildly-bound outflow with specific angular momentum somewhat lower than that of corotation at $L_2$ (Figures \ref{fig:bounddist} and \ref{fig:boundstream}).

\item Unbound ejecta are generated only very late in the binary coalescence, as the binary separation becomes similar to the radius of the donor.  By this time the orbital motion is desynchronized from the donor's rotation.  A broad fan of ejecta forms (Figure \ref{fig:binaryflow1}) as the accretor skims the surface of the donor.  With the gravitational slingshot from the secondary, much of the envelope gas reaches high enough radial velocity to become unbound just after passing the accretor (Figures \ref{fig:unbounddist} and \ref{fig:unboundstream}). 

\item As unbound ejecta interact with the earlier, bound outflow, they are decelerated in the binary midplane and redirected toward the poles (Figure \ref{fig:Evertical}). The morphology that emerges is a dense, bound torus of tens of donor-star radii, with lower-density but higher-entropy and higher-velocity unbound material emerging along broad (roughly 40 degree) bipolar cones (Figure \ref{fig:azimuthalavg}). 

\end{enumerate}

As we discuss, we believe our  findings are directly applicable to the morphologies observed or inferred in remnants of galactic red novae. 
More speculatively, our findings may be broadly related to the morphology of bipolar planetary nebulae as discussed by \citet{2018ApJ...860...19G} and \citet{2018arXiv180705925F}, particularly as some planetary nebulae seem likely to be remnants of common-envelope interactions, as reviewed in detail by \citet{2017PASA...34....1D,2017NatAs...1E.117J}. Certainly the discovery of similarly bipolar morphologies as revealed in pre-planetary nebulae by ALMA \citep[e.g.][]{ 2017ApJ...841..110S,2018arXiv180401610K} and a  universal potential explanation for the origin of this feature, as presented in this work, make this connection all the more appealing. 

The work presented here is, however, far from exhaustive. We have tested that our qualitative conclusions are robust to changes in binary mass ratio and initial spin synchronization, and we have presented results for a mass ratio, $q=0.3$, previously predicted to produce among the most-unbound early outflows \citep[see][Figure 3]{2016MNRAS.455.4351P}. However, future work should include a consideration of a broader set of binary parameters and studies of how these affect the emergent distributions of material. The microphysics of gas ionization state, opacity, and cooling likely all play important roles in determining the scale height of the circumbinary structure \citep{2016MNRAS.455.4351P,2016MNRAS.461.2527P} and perhaps even the details of mass loss from the accretor's Roche lobe. 

Our simulation  terminates with the engulfment of the accretor.
Following simulated systems further into the the common-envelope interaction could better address how the circumbinary gas distribution interacts and shapes subsequently-disturbed material. In cases in which the binary eventually merges, as is thought to be the case with the galactic luminous red novae transients, winds from the young merger remnant could be dynamically significant. In cases where a binary survives, interaction between multiple components of outflow and ejecta may continue to be important as the binary slowly clears its surroundings. 

\acknowledgements
This work would not have been possible without many informative conversations with T. Kaminski. We also thank the participants of the ``Evolved stars and binaries'' group at the CfA, Jonathan Grindlay, Abraham Loeb, Enrico Ramirez-Ruiz, and Rosanne Di Stefano for many helpful discussions. 
We gratefully acknowledge collaboration with Wenrui Xu on the implementation of the pole-averaging algorithm used here. 
M.M. is grateful for support for this work provided by NASA through Einstein Postdoctoral Fellowship grant number PF6-170169 awarded by the Chandra X-ray Center, which is operated by the Smithsonian Astrophysical Observatory for NASA under contract NAS8-03060. 
Support for program \#14574 was provided by NASA through a grant from the Space Telescope Science Institute, which is operated by the Association of Universities for Research in Astronomy, Inc., under NASA contract NAS 5-26555.
The work of E.C.O. was supported by a grant from the Simons Foundation (grant no. 510940). 
Resources supporting this work were provided by the NASA High-End Computing (HEC) Program through the NASA Advanced Supercomputing (NAS) Division at Ames Research Center.

\software{ Astropy \citep{2013A&A...558A..33A}, Athena++, Stone et al. (in preparation) \url{http://princetonuniversity.github.io/athena} }

\bibliographystyle{aasjournal}

\end{document}